\documentclass{article}
\usepackage[utf8]{inputenc}
\usepackage[a4paper,margin=2.5cm]{geometry}
\usepackage{amsmath,amssymb}
\usepackage{booktabs}
\usepackage{graphicx}
\usepackage{subcaption}
\usepackage{xcolor}
\usepackage{url}
\usepackage[normalem]{ulem}

\newcommand{\Lop}{\hat{\mathrm{L}}}

\title{Optical symmetry breaking for single-image field reconstruction}
\newcommand{\articleAuthor}{Egor Manuylovich}
\author{\articleAuthor\\
\small Aston Institute of Photonic Technologies, Aston University, Birmingham B4 7ET, UK\\
\small \texttt{e.manuylovich@aston.ac.uk}}
\date{}

\usepackage{tabularx,xurl,siunitx,float,placeins}

\begin{document}
\makeatletter
\let\arxivArticleTitle\@title
\makeatother
\let\arxivAddContentsLine\addcontentsline
\renewcommand{\addcontentsline}[3]{}
\maketitle
\begin{abstract}
Light carries information in both amplitude and phase, although cameras record only intensity. Reconstructing the optical field is at the heart of wavefront sensing, microscopy and coherent communications. Existing methods often rely on multiple measurements or iterative phase retrieval. In fields described by known spatial modes, intensity is determined by pairwise products of their complex coefficients. Yet optical symmetries can make these products indistinguishable, leaving different fields with nearly identical images. Here, we show that engineering the optical measurement to break these symmetries enables direct reconstruction of the complex field from a single intensity image. We optimise a phase mask followed by propagation to separate pairwise modal signatures while retaining light on the detector. A calibrated linear inverse and rank-one projection then recover modal amplitudes and relative phases without iterative phase retrieval. In simulations, optimised masks satisfy conditioning and light-capture requirements unmet by random masks. Experiments reconstruct 42-mode fibre fields using either a spatial light modulator or a passive diffuser, with median complex-field errors of 9.1\% and 4.6\%, respectively, against holographic references. This extends an earlier five-mode experimental demonstration of fast intensity-only modal decomposition. Per-frame computation takes 2.1\,ms, and batched throughput exceeds 4,000 frames per second. These results establish measurement-space symmetry breaking as a route to rapid single-image optical field reconstruction.
\end{abstract}

Optical detectors measure intensity, while a complete description of a coherent light field requires both amplitude and phase. Recovering the complex field from intensity measurements is important for wavefront sensing \cite{norris2020all}, quantitative label-free microscopy \cite{cao2024high}, astronomical adaptive optics \cite{norris2020all}, optical coherence tomography \cite{badon2016smart}, multimode fibre imaging \cite{wen2023single}, and optical metrology \cite{zuo2022deep}. Conventional phase-retrieval methods infer the missing phase information by iteratively enforcing consistency between a candidate field, measured intensities and known physical constraints \cite{fienup1982phase,shechtman2015phase}. Yet their success is fundamentally limited by the information encoded in the optical measurement itself: if distinct fields produce identical intensity patterns, no reconstruction algorithm can distinguish between them.

We refer to this loss of distinguishability as \emph{measurement degeneracy}. Exact degeneracy occurs when distinct fields produce identical measurements, while near-degeneracy makes their reconstruction highly sensitive to noise \cite{grohs2020phase}. For example, intensity measured directly in the field plane cannot distinguish a field from its complex conjugate, while Fourier-intensity measurements admit a related conjugate-inversion ambiguity \cite{candes2015coded}. These ambiguities go beyond the unavoidable global-phase uncertainty inherent to intensity-only measurements: they prevent the unique determination of the relative phases that define the field. Reliable phase reconstruction therefore requires not only an effective inversion algorithm, but also an optical measurement that makes these relative phases observable.

One way to overcome these limitations is to engineer the optical measurement so it allows to observe otherwise unobservable phase information. Phase diversity achieves this by introducing known aberrations across multiple images \cite{gonsalves1982diversity}, while coded diffraction uses controlled modulations to provide additional intensity constraints \cite{candes2015coded}. In wavefront sensing, the asymmetric-pupil Fourier sensor exploits pupil asymmetry to enable linearised phase recovery in the small-aberration regime \cite{martinache2013asymmetric}. More recently, computational wavefront sensors have extended this principle to single-shot measurements, using a diffuser or random phase plate to encode the complex field into a single intensity image \cite{gao2026single}. However, reconstructing the field still requires solving a regularised, nonconvex optimisation problem, typically involving hundreds of iterations, parameter tuning and seconds of computation per frame. Although capable of recovering arbitrary fields at high spatial resolution, these methods do not guarantee unique reconstruction \cite{gao2026single}.

For fields constrained to a known modal basis, such as those in multimode fibres and waveguides, optical phase reconstruction can be expressed as recovery of complex modal coefficients. Their powers describe the distribution of energy among modes, while their relative phases determine interference and the resulting wavefront. Measuring both supports the characterisation of fibre-laser dynamics \cite{stutzki2011high}, analysis of mode excitation and coupling \cite{flamm2013digital}, and wavefront sensing through modal mappings \cite{norris2020all}.

Existing approaches to modal decomposition differ in the optical measurements they require and how they extract modal information. Spatially and spectrally resolved imaging distinguishes modes through their spectral interference signatures \cite{nicholson2008spatially,nguyen2012modal}, whereas cross-correlated imaging uses an external reference to recover modal profiles, powers and delays \cite{schimpf2011cross,demas2014subsecond}. Computer-generated holographic correlation filters provide another way by measuring modal projections and interference terms to determine complex modal coefficients \cite{kaiser2009complete,flamm2012mode}, with experimental demonstrations involving approximately 80 fibre modes \cite{gervaziev2021mode}. Alternatively, computational methods infer modal coefficients directly from intensity measurements, using fibre-specific iterative optimisation \cite{lu2013fast,li2017hybrid}, machine-learned approaches \cite{an2019learning,rothe2020deep}, or neural-network estimates followed by numerical refinement \cite{kim2023convolutional}. Additional modal information can also be obtained by introducing measurement diversity through fractional-Fourier transformations \cite{yan2019modal}. More recently, coherent modulation imaging has demonstrated single-exposure, reference-free modal reconstruction in few-mode fibres, although recovering the field still involves computational phase retrieval and simulated-annealing mode projection \cite{zou2025reference}.

A different approach exploits the fact that the intensity of a modal superposition can be expressed as a linear combination of pairwise mode products (PMPs), weighted by products of the complex modal coefficients \cite{manuylovich2020fast,manuylovich2021intensity}.  This formulation enables modal reconstruction through linear inversion, provided that the measured PMP signatures contain sufficient independent information. Sparse intensity sampling can reduce the number of measurements by selecting informative detector positions within a given intensity basis \cite{manuylovich2025sparse}. However, selecting where to measure cannot resolve information that is fundamentally absent from the optical measurement itself. Symmetries in the modal basis and observation geometry can produce identical or nearly linearly dependent PMP signatures, making their inversion ambiguous or sensitive to noise. Optical measurement design therefore addresses a more fundamental question: how can the optical transformation before detection be engineered to make these signatures distinguishable?

In this work, we introduce an optical symmetry-breaking approach that improves the observability of coherent fields from intensity-only measurements. By applying a phase transformation followed by propagation before detection, we make otherwise overlapping pairwise mode-product (PMP) signatures more distinguishable without altering the original modal basis. We optimise this transformation to improve PMP conditioning while maintaining sufficient power capture, with the resulting condition number also providing a bound on linear noise amplification. In simulations, a phase mask optimised for 42 modes enables reconstruction across fibres supporting 6--51 modes using fibre-specific inverses. Experimentally, we demonstrate single-frame intensity-only decomposition of 42-mode fibre fields using either a spatial light modulator or a passive ground-glass diffuser, increasing the experimentally accessible mode count for fast decomposition methods from five to 42, with scope for further scaling to higher-mode-count fibres. Together, these results establish an order-of-magnitude increase in experimentally accessible mode count over previous fast intensity-only decomposition methods. We validate the reconstructed fields against held-out off-axis holographic measurements, achieving a single-frame processing latency of 2.1 ms and batched throughput exceeding 4,000 frames per second on a consumer-grade GPU. 

By breaking optical symmetries in the measurement space, the proposed approach improves field observability independently of the reconstruction algorithm, potentially benefiting not only fast mode decomposition but also neural-network-based and other computational wavefront reconstruction methods.

\section{Results}

For a field constrained to $M$ known  modes $\Psi_j$, the complex field in the fibre-facet (near-field) plane $z_{\mathrm{nf}}$ is

\begin{equation}
    E(x,y,z_{\mathrm{nf}})=\sum_{j=1}^{M} C_j\Psi_j(x,y,z_{\mathrm{nf}}),
    \label{eq:field-expansion}
\end{equation}

where $C_j$ are complex modal coefficients. A linear optical operator $\Lop$ maps each mode to the camera plane, and the measured intensity is

\begin{equation}
    I(x,y)=\sum_{j=1}^{M}\sum_{k=1}^{M}
    C_j\overline{C_k}\,
    \Lop[\Psi_j](x,y)\overline{\Lop[\Psi_k](x,y)}.
    \label{eq:intensity-expansion}
\end{equation}

\begin{figure}[!ht]
  \centering
  \includegraphics[width=\linewidth]{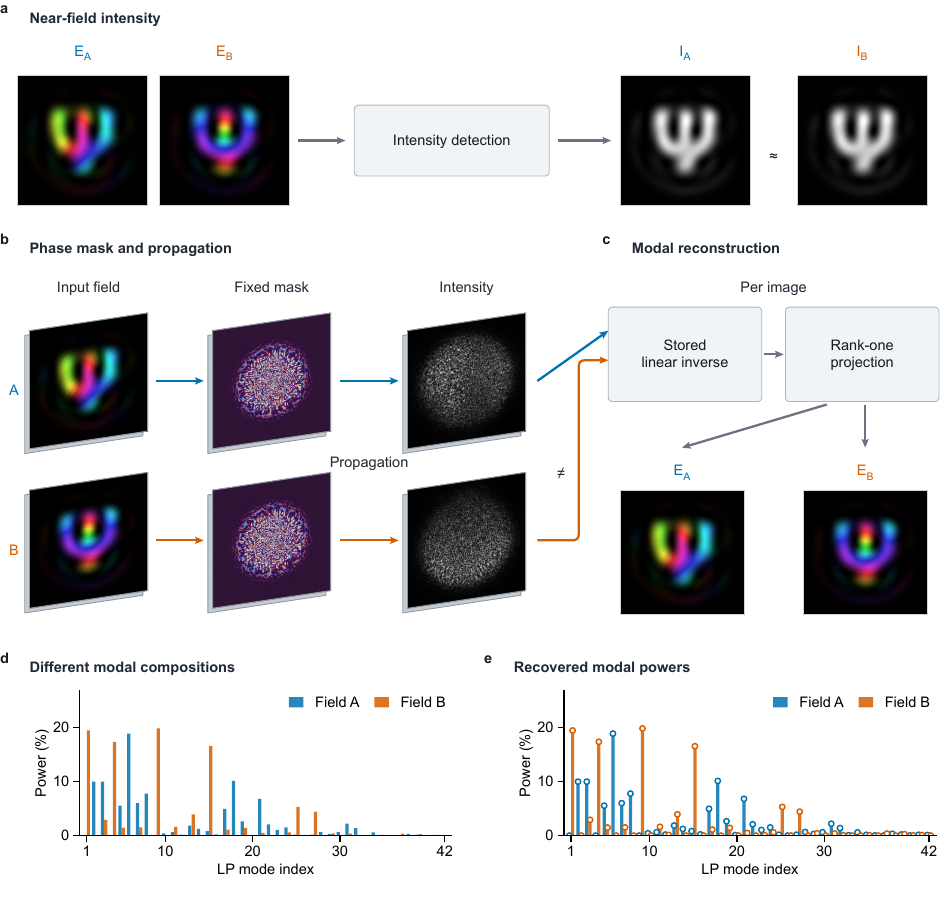}
\includegraphics[width=0.28\linewidth]{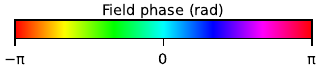}\quad\includegraphics[width=0.28\linewidth]{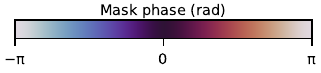}
  \caption{Measurement-space symmetry breaking enables direct modal phase reconstruction.
  \textbf{(a)} Two fields of the 42-mode fibre, $E_A$ and $E_B$, have near-field intensities with normalised overlap 0.99998 despite different modal compositions.
  \textbf{(b)} A fixed phase mask followed by free-space propagation in the simulated optical system maps the same two fields to distinct camera intensities (normalised overlap 0.42). The mask is an L1-regularised version of the optimised mask used for the simulations of Fig.~\ref{fig:direct-statistics} and displayed on the SLM in the experiment of Fig.~\ref{fig:experimental-examples}, with its phase modulation confined to a disc of 2.3 mm radius (Methods).
  \textbf{(c)} Because the PMP matrix is well conditioned in the symmetry-broken measurement space, each intensity image is multiplied by its pseudoinverse to recover the pairwise products of the modal weights unambiguously. A rank-one projection of these products gives the modal weights, which synthesise the field in the original basis. Reconstructed fields are globally phase-aligned for display.
  \textbf{(d)} Normalised modal powers, $|C_j|^2/\sum_k|C_k|^2$, reveal the different modal structures.
  \textbf{(e)} Recovered powers (open circles) agree with the corresponding true powers (bars). Modes are ordered by azimuthal index $l$ and then radial index $m$, with the two angular partners of each LP$_{lm}$ mode with $l>0$ adjacent. Hue encodes phase and brightness intensity; greyscale shows linear near-field intensity and the square root of the camera intensity (central $4\times4$ mm), with shared scales within each plane. Noiseless simulation (Methods).}
  \label{fig:measurement-symmetry-breaking}
\end{figure}

The observable structures are therefore not the modes alone but their pairwise products. Under conventional near- or far-field measurements, symmetry can make several products identical or nearly dependent. Figure~\ref{fig:measurement-symmetry-breaking} illustrates this ambiguity with two fields of the simulated 42-mode fibre: their modal powers and relative phases differ, but their near-field intensities have a normalised overlap of 0.99998. Applying the same phase transformation and propagation to each field produces distinct intensity signatures, from which each field is recovered in the original basis.

After sampling the camera at $N$ pixels and flattening each pattern, Eq.~\ref{eq:intensity-expansion} becomes
\begin{equation}
    \mathbf{I}=\mathbf{\Upsilon}(\Lop)\mathbf{S},
    \qquad
    S_{M(j-1)+k}=C_j\overline{C_k},
    \label{eq:pmp-system}
\end{equation}
where the columns of the PMP matrix $\mathbf{\Upsilon}(\Lop)\in\mathbb{C}^{N\times M^2}$ are the sampled pairwise products in the measurement plane,
\begin{equation}
    \boldsymbol{\upsilon}_{M(j-1)+k}(\Lop)=
    \Lop[\Psi_j]\circ\overline{\Lop[\Psi_k]}.
    \label{eq:pmp-column}
\end{equation}
We use its condition number, $\mathcal{D}(\Lop)=\kappa[\mathbf{\Upsilon}(\Lop)]$, as a quantitative measure of measurement degeneracy: a large value indicates nearly dependent observable patterns and correspondingly strong noise amplification on inversion. A well-conditioned PMP matrix is a sufficient route to recovering all coefficient products; it is not necessary for uniqueness on the smaller set of physically admissible rank-one products. Near- and far-field measurements are simply two choices of $\Lop$, the identity and the Fourier transform. The ten-mode example has a singular near-field matrix and a poorly conditioned far-field matrix (Fig.~\ref{fig:pmp-conditioning-similarity}a; Supplementary Note~1). The condition number of such conventional measurements grows rapidly with the number of modes \cite{manuylovich2021intensity}.

\subsection{Engineering the measurement operator}

We implement a measurement operator $\Lop_{\Phi}$ comprising propagation from the fibre to a lens, propagation to a spatial phase mask $\Phi$, and propagation to a camera. The mask can be programmed on an SLM during development and later realised as a static phase plate, a two-photon-polymerised structure or a metasurface. $\Lop_{\Phi}$ does not redefine the modal basis in Eq.~\ref{eq:field-expansion}: it changes the detector signatures of that fixed basis, and therefore its observability, while the mode count and coefficients remain the same.

The propagation model is differentiable with respect to the mask pixels. For a given mask, the propagated basis modes $E_j(\mathbf{r};\Phi)=\Lop_{\Phi}[\Psi_j](\mathbf{r})$ form the PMP matrix $\mathbf{\Upsilon}(\Lop_{\Phi})$. We minimise its condition number subject to keeping the light on a detector region of interest (ROI),
\begin{equation}
    \Phi_{\mathrm{opt}} = \underset{\Phi}{\operatorname{argmin}}\;
    \kappa\!\left[\mathbf{\Upsilon}(\Lop_{\Phi})\right]
    \quad\text{subject to}\quad
    \bar\eta(\Phi)\geq\eta_{\mathrm{mean}},\qquad
    \eta_{\min}(\Phi)\geq\eta_{\mathrm{worst}},
    \label{eq:optimization}
\end{equation}
where $\bar\eta$ is the fraction of each basis mode's power captured in the ROI, averaged over modes, and $\eta_{\min}$ is the smallest captured fraction over every coherent superposition of the modes. Minimising $\kappa$ promotes distinguishable detector signatures for the fixed basis, while the second constraint guarantees capture for any field in the modal subspace, not only for the individual modes. The parameterisation, constraints and optimisation settings are given in Methods and Supplementary Notes~3 and 4.

\begin{figure}[!ht]
  \centering
  \includegraphics[width=0.6\linewidth]{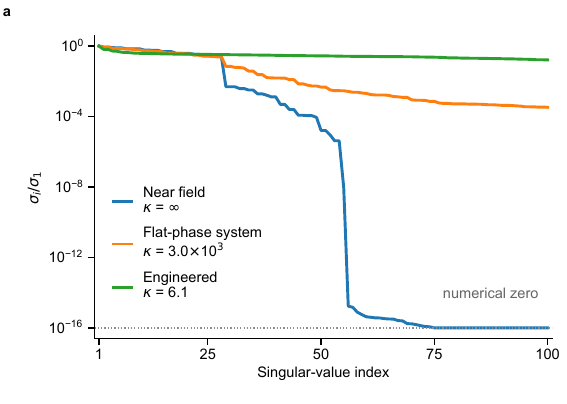}\hfill
  \includegraphics[width=0.3875\linewidth]{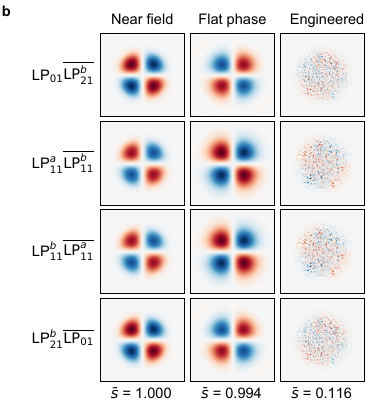}
  \caption{Symmetry breaking removes measurement degeneracy. Ten-mode LP basis (core radius 6\,$\mu$m, numerical aperture 0.10; 100 pairwise products) in the simulated optical system.
  \textbf{(a)} Normalised singular values of the near-field, flat-phase and engineered PMP matrices, with their condition numbers $\kappa$. The flat-phase result uses the same optical train with zero SLM phase; the engineered measurement uses the optimised mask on the SLM. Values below $10^{-16}$ are drawn at that level.
  \textbf{(b)} A group of near-degenerate products. Each row shows the real part of a mode product in the near field, the flat-phase system and the engineered measurement; $\bar s$ is the mean absolute normalised overlap of the full complex products. Conjugate products have identical real parts but opposite imaginary parts. The full complex products are nearly dependent before encoding and become distinguishable after the engineered transformation. Selected near-field degeneracies of the 42-mode fibre are shown in Supplementary Fig.~1; its 29 four-member groups are listed in Supplementary Table~1.}
  \label{fig:pmp-conditioning-similarity}
\end{figure}

Figure~\ref{fig:pmp-conditioning-similarity} compares the resulting PMP matrix with those of near-field and flat-phase measurements. The engineered system has a much flatter singular-value spectrum and a lower condition number, so all measurement directions are represented more uniformly and individual product signatures become more distinguishable. The product patterns themselves show why (Fig.~\ref{fig:pmp-conditioning-similarity}b): in the near field and flat-phase system, different pairs of LP modes generate identical or nearly dependent signatures, and these near-degenerate directions are precisely those that limit the conditioning. After the optimised phase transformation they become distinct.

\begin{figure}[!ht]
\centering
\includegraphics[width=\linewidth]{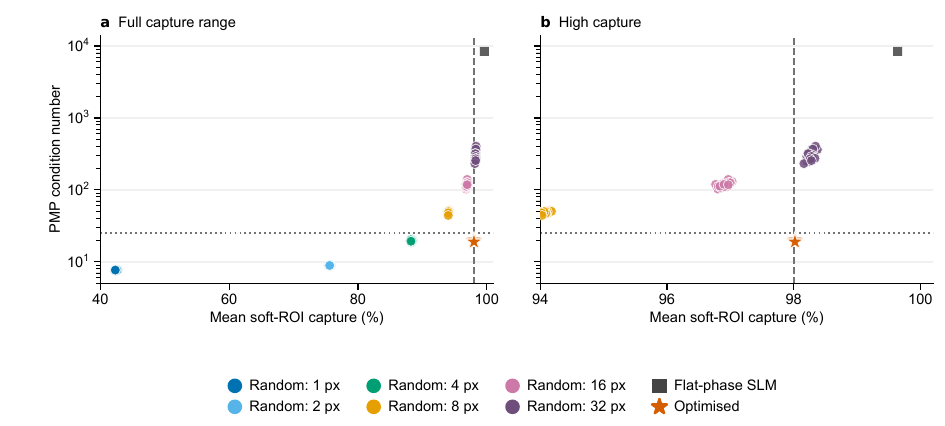}
\caption{Optimised masks combine low condition number with high power capture. Full-camera PMP condition number versus mean soft-ROI power capture for 42 modes: 144 tiled random masks with cell sizes of 1 to 32 pixels (24 per size), 12 optimised masks and a flat-phase SLM. The 72 smooth random controls are reported separately in Supplementary Note~4. Soft-ROI capture is spatially weighted power capture. (b) enlarges the high-capture region. The dashed and dotted lines mark the 98\% capture and condition-25 targets; a second constraint, 95\% capture for every coherent superposition, is evaluated separately. All systems share the same optics, phase quantisation and incident-power normalisation.}
\label{fig:roi-mask-controls}
\end{figure}

A random phase mask can break measurement degeneracies, but good conditioning alone does not keep the light within a useful detector footprint. In a matched 42-mode simulation we therefore compared a flat-phase SLM in the same optical system, 216 random masks spanning spatial granularity and smoothness, and 12 solutions of Eq.~\ref{eq:optimization} from independent initial phases, with $\eta_{\mathrm{mean}}=98\%$ and $\eta_{\mathrm{worst}}=95\%$ for a soft ROI of radius 2 mm (Fig.~\ref{fig:roi-mask-controls} and Methods).

All 12 optimised masks meet both capture limits with a condition number below 25 (median 19.3), whereas none of the 216 random masks meets this joint specification. Optimisation also reduces peak detector loading without targeting it explicitly, improving the joint trade-off between conditioning, capture and loading (Supplementary Note~4).

\subsection{Single-image reconstruction}

Once the measurement operator is fixed and the camera-plane modal fields are known, reconstruction proceeds through the PMP matrix $\mathbf{\Upsilon}$. For a measured intensity image $\mathbf I$, the coefficient products are estimated by
\begin{equation}
\mathbf S
=\mathbf{\Upsilon}^{\dagger}\mathbf I,
\label{eq:linear-reconstruction}
\end{equation}
so inference involves no iterative fitting of the unknown field.

The recovered product vector is reshaped into an $M\times M$ Hermitian matrix $\widehat{\mathbf S}$. For an ideal coherent field this matrix equals $\mathbf C\mathbf C^{\mathrm H}$ and is positive semidefinite and rank one; noise and model mismatch violate these properties. We therefore retain the leading eigenpair $(\mu_1,\mathbf v_1)$,
\begin{equation}
\widehat{\mathbf S}_1=\mu_1 \mathbf v_1\mathbf v_1^{\mathrm H},
\qquad
\widehat{\mathbf C}=\sqrt{\mu_1}\,\mathbf v_1,
\label{eq:coefficient-reconstruction}
\end{equation}
which gives the modal amplitudes and relative phases up to the unobservable global phase, and synthesise the near field in the original basis,
\begin{equation}
\widehat E(x,y,z_{\mathrm{nf}})=\sum_{j=1}^{M}\widehat C_j\Psi_j(x,y,z_{\mathrm{nf}}).
\label{eq:reconstructed-field}
\end{equation}
Only the measured intensity enters reconstruction, and the computation per image is one matrix--vector product with the stored inverse followed by an $M\times M$ eigendecomposition; the inference latency and throughput are reported below.

\begin{figure}[!ht]
\centering
\includegraphics[width=\linewidth]{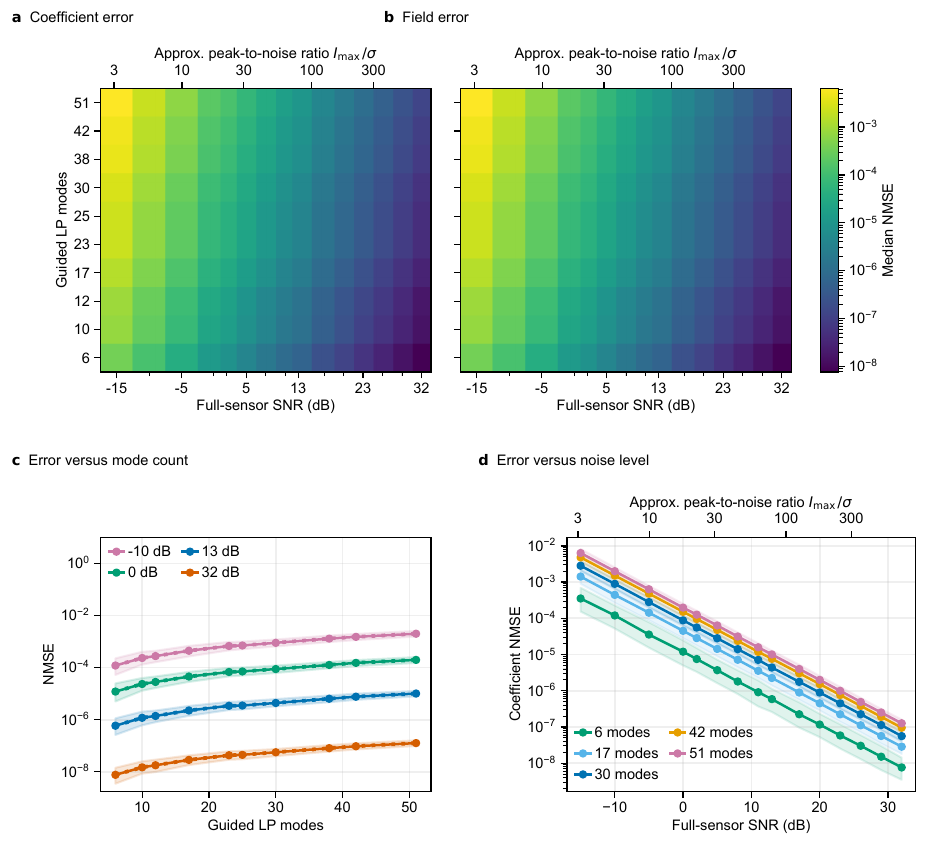}
\caption{One mask designed for 42 modes supports single-image reconstruction from 6 to 51 modes. The mask and optics are fixed while the core radius and noise level vary. Targets below 5 dB use unclipped Gaussian noise (Methods). Bottom axes give the full-sensor SNR (Methods); top axes give the equivalent peak-to-noise ratio $I_{\max}/\sigma$, the brightest pixel over the standard deviation of the added noise, placed with the median offset of 24.9 dB between the two scales (Methods). (a,b) Median coefficient and complex-field NMSE over 1,000 independent field/noise trials per cell, at 15 SNR levels from $-15$ to 32 dB. (c) Error versus mode count at four SNRs; dotted curves are field errors. (d) Coefficient error versus measured SNR for selected mode counts. Shading shows empirical 5th--95th percentiles of coefficient error, not confidence intervals. The same fields are paired across noise levels; all trials and outliers are retained. One least-squares inverse is used per radius for all noise levels (Methods). A median-error example and full distributions are given in Supplementary Note~5.}
\label{fig:direct-statistics}
\end{figure}

To test transfer of the optical design, we kept the 42-mode-optimised mask and the optics fixed and varied the fibre core radius from 4.1 to 14\,$\mu$m (numerical aperture 0.10, wavelength 632.8 nm), giving 6 to 51 guided LP modes, and the full-sensor intensity signal-to-noise ratio (SNR) from $-15$ to 32 dB, with 1,000 random fields per condition (Fig.~\ref{fig:direct-statistics} and Methods). A separate inverse was constructed for each fibre radius. The NMSE is approximately inversely proportional to the linear signal-to-noise ratio over the tested range and grows smoothly with mode count. At 13 dB ($I_{\max}/\sigma\approx80$), the median coefficient normalised mean-squared error (NMSE) is $7.6\times10^{-6}$ for 42 modes and $1.0\times10^{-5}$ for 51 modes, corresponding to relative $L^2$ errors of 0.28\% and 0.32\%, and the complex-field errors closely follow. The low errors reflect averaging of independent noise over the full sensor under a matched optical model; extreme-noise results and their limits are detailed in Supplementary Note~5.

\subsection{Experimental single-image reconstruction}

\begin{figure}[!ht]
\centering
\includegraphics[width=\linewidth]{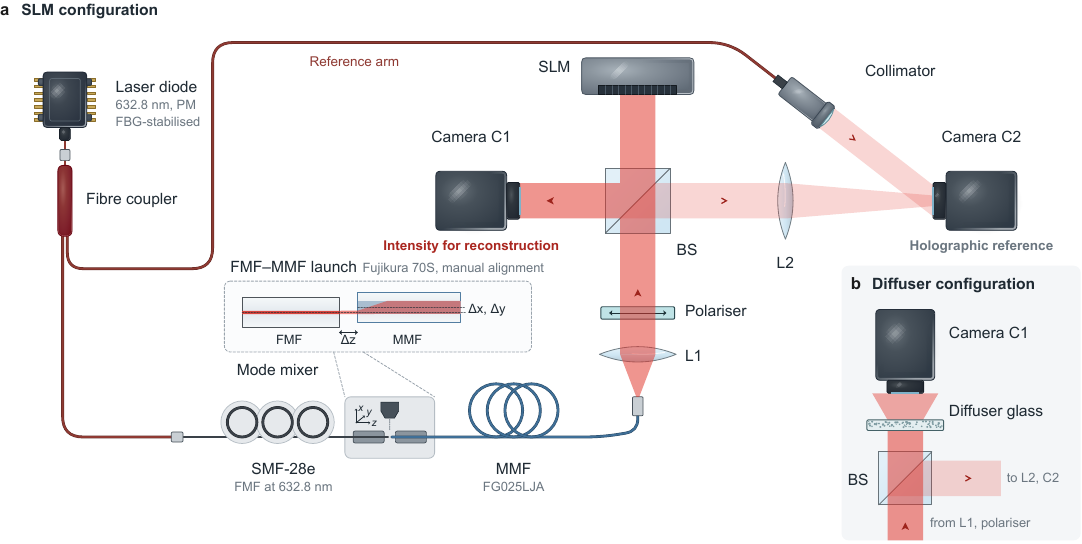}
\caption{Experimental setup. \textbf{(a)} SLM configuration. A fibre coupler divides the output of a 632.8 nm FBG-stabilised laser diode into signal and reference arms. In the signal arm, the field in few-mode fibre (FMF; SMF-28e) passes a motorised polarisation controller used as a mode mixer and is launched into multimode fibre (MMF) across a free-space gap (inset). The fibres are not spliced: they are held in the V-groove clamps of a Fujikura 70S in manual-alignment mode, whose three-axis ($x$, $y$, $z$) positioning and microscope set and measure the transverse core offset ($\Delta x$, $\Delta y$) and the axial gap ($\Delta z$). The collimated (L1) and polarised MMF output is divided by a beamsplitter (BS). The transmitted field returns from the reflective SLM and propagates to camera C1, which records the symmetry-broken (symmetry-broken-plane) intensity used for reconstruction. Lenses L1 and L2 form a 4-f imaging system that projects the fibre near field, reflected by the BS, onto camera C2 for off-axis digital holography using a collimated reference beam. \textbf{(b)} Diffuser configuration: the SLM is removed and the transmitted field passes a ground-glass diffuser directly onto C1, which faces the BS; all other components are as in (a). Strong red marks the path to C1, the only reconstruction input; pale red marks the holographic paths to C2. Spacings, beam sizes, the launch gap, offset and component models are given in Methods.}
\label{fig:experimental-setup}
\end{figure}

We implemented the engineered measurement with either a reflective SLM displaying the optimised mask (Methods), or a ground-glass diffuser in front of the reconstruction camera (Fig.~\ref{fig:experimental-setup} and Methods). A few-mode fibre is launched into a multimode fibre with a controlled transverse offset and axial gap, and a polarisation controller varies the excited superposition, which we represent by 42 LP modes. For each field, camera C1 records the symmetry-broken-plane intensity, and camera C2 records an off-axis hologram (and, in the diffuser acquisition, also the near-field intensity). Each acquisition comprises 5,000 such measurements from 250 launch conditions with 20 controller states each. We held out 25 launch conditions (500 fields), calibrated the camera-plane modal fields from the holographic labels of the remainder and held the calibrated camera-plane modal fields fixed. Each held-out field was then reconstructed from its symmetry-broken-plane image alone and compared with its hologram.

\begin{figure}[!tbp]
\centering
\begin{minipage}[t]{0.57\linewidth}\vspace{0pt}%
  \includegraphics[width=\linewidth]{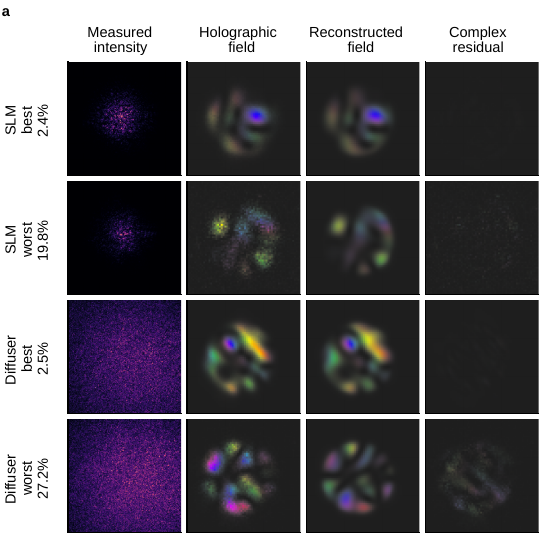}%
\end{minipage}\hfill
\begin{minipage}[t]{0.41\linewidth}\vspace{0pt}%
  \textbf{b}\\\includegraphics[width=\linewidth]{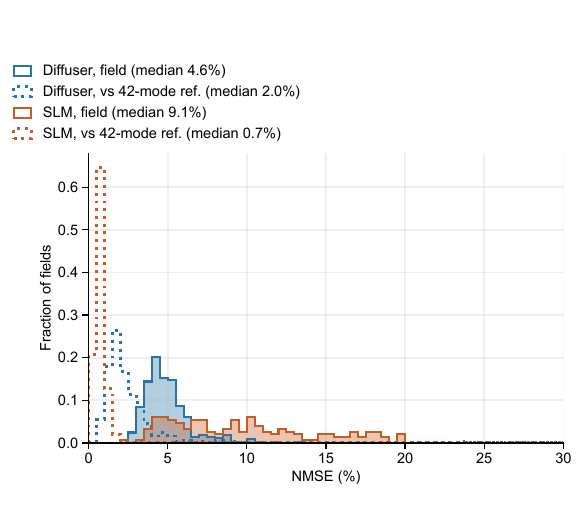}%
\end{minipage}\\[4pt]
\includegraphics[width=\linewidth]{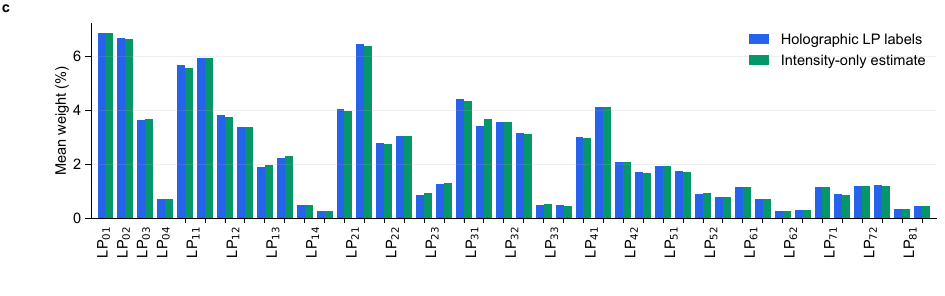}
\includegraphics[width=0.28\linewidth]{figures/display_keys/field_phase.pdf}
\caption{Single-image reconstruction of 42-mode fibre fields in experiment. \textbf{(a)} Best and worst eligible test fields for the SLM displaying the optimised mask (top two rows) and for the diffuser (bottom two rows): measured symmetry-broken-plane intensity on C1, holographic reference field, field reconstructed from the intensity alone, and complex residual; the percentage beside each row is its complex-field NMSE. Reconstructions are aligned by one complex scalar and share the reference intensity scale within each row; hue encodes phase, and brightness and saturation encode intensity. These are selected extremes. The worst SLM example has 1.0\% NMSE relative to the 42-mode projection of its reference; most of its full-reference discrepancy lies outside that subspace. \textbf{(b)} Full-reference field NMSE distributions (filled) and NMSE relative to the LP-projected reference (dotted), with 368 diffuser and 150 SLM fields; each distribution is normalised by its sample count. The calibrated PMP spectra are shown in Supplementary Fig.~13. \textbf{(c)} Mean normalised coefficient weights of all 42 LP components of the 150 eligible SLM test fields, from the holographic references (blue) and from the symmetry-broken-plane intensity alone (green); the two angular partners of each LP$_{lm}$ mode with $l>0$ are adjacent ($a$ first). Each field and method is normalised to unit total weight before averaging. Weights are in the fitted LP basis, not calibrated optical powers. Error distributions, examples of excluded references, further examples and the complete metrics are given in Supplementary Note~6 and Supplementary Table~9.}
\label{fig:experimental-examples}
\end{figure}

The calibrated PMP matrices have condition numbers of approximately 107 for the diffuser and 45 for the SLM; all 1,764 product directions are retained (Supplementary Fig.~13). Holograms containing saturated pixels or failing reference-quality checks cannot verify a reconstruction and were excluded before calibration, leaving 368 of 500 diffuser and 150 of 500 SLM test fields (Methods). Eligibility concerns the holographic reference rather than the reconstruction channel (Supplementary Note~6). For the eligible fields, the median complex-field NMSE after global complex-scalar alignment is 4.6\% for the diffuser and 9.1\% for the SLM, with median field coherences of 0.977 and 0.953 (Fig.~\ref{fig:experimental-examples}a and Supplementary Fig.~13).

The field error includes noise and out-of-subspace light in the holographic reference. Against its 42-mode projection, the median reconstruction NMSE is 0.7\% for the SLM and 2.0\% for the diffuser (Supplementary Note~6 and Supplementary Table~9). The recovered LP weights agree with the holographic labels (Fig.~\ref{fig:experimental-examples}c). The diffuser result establishes that a calibrated passive scattering element suffices for single-image reconstruction.

Near-field intensity alone fails with the same inverse, rank-one projection and LP basis: only 286 of the 1,764 product directions are retained. The median field NMSE is 94\% for measured reference-blocked diffuser images, and 96\% (diffuser) or 98\% (SLM) for intensities computed from the holographic references (Supplementary Note~6).

Each image is reconstructed without iterative phase retrieval. On a consumer graphics processor (NVIDIA RTX 3090), the complete pipeline from a raw $1,024\times1,280$ camera frame to the 42 complex modal weights, comprising preprocessing, one product with the stored $1,764\times65,536$ regularised pseudoinverse and the $42\times42$ eigendecomposition, takes a median of 2.1 ms per frame for both operators when frames are processed one at a time, that is about 480 frames per second. Processing frames in batches raises the throughput to 4,000--4,200 frames per second for frames held in GPU memory, and to about 2,100 frames per second including their transfer from host memory (Methods and Supplementary Note~5). Both cameras were operated at 60 frames per second, C1 (Ximea) over its full sensor and C2 (Thorlabs) over a $512\times512$-pixel region. The single-frame latency is below the 16.7 ms acquisition period.

The two acquisitions differ in launch conditions, modal content, reference quality and calibration, so they do not rank the two elements: the SLM has the lower condition number and the lower error relative to the 42-mode reference, but the larger field error, because the field error also reflects the quality of each reference.

\section{Discussion}

Measurement-space symmetry breaking improves field observability by making distinct field states produce more distinguishable intensity images. The PMP condition number quantifies the noise amplification factor and robustness of this approach. Greater distinguishability in the recorded images could also improve performance  of other mode decomposition methods, such as neural networks trained on the encoded measurements and for iterative wavefront reconstruction methods that incorporate the optical transformation in their forward model.

Optimisation is also distinct from the mere presence of a mask. A generic asymmetric mask can break some degeneracies, but it need not distribute information efficiently over a finite, noisy and pixelated sensor; in the matched simulation, only optimised masks combined good conditioning with high power capture and low peak loading.

The phase mask can be static after design and calibration, while the procedure is not specific to fibre modes. Applying the same optimisation to 36 Hermite--Gaussian beams and 41 square-waveguide modes gives $\kappa=13.5$--14.2 and 17.5--18.8 over three initial phases, at 98\% mean capture. At 13 dB, median coefficient NMSEs are $5\times10^{-6}$ and $7\times10^{-6}$. In both families the near- and far-field matrices are singular, the flat-phase SLM has $\kappa\approx8\times10^3$--$10^4$, and no tested random mask meets the joint conditioning and capture targets. The fibre-optimised mask transfers unchanged, giving $\kappa=16.8$ and 24.1; re-optimisation improves both (Supplementary Note~7, Supplementary Table~10 and Supplementary Fig.~20).

\section{Methods}

\subsection{Forward model and calibration}

For each known input mode, the field is propagated through the same optical scheme. The sampled camera-plane fields form the PMP columns in Eq.~\ref{eq:pmp-column}. In simulation these fields are computed with scalar Fresnel propagation. In the experiments, these fields were calibrated from holography-derived labels, incorporating repeatable system aberrations without requiring an exact component-level model. The numerical parameters are listed in Supplementary Table~3.

\subsection{Experimental scheme and modal excitation}

Fig.~\ref{fig:experimental-setup} shows the SLM (a) and diffuser (b) configurations. The source was a fibre-Bragg-grating-stabilised laser diode (PL-NL-0633-BA81-PP-PZT with wavelength of 633 nm, linewidth <1 MHz, PM630-pigtailed). A fibre coupler divided its output into signal and reference arms. In the signal arm, SMF-28e fibre, which guides 10 modes at this wavelength, served as the few-mode fibre (FMF). A motorised polarisation controller (Thorlabs MPC320) acted as a mode mixer, its settings are the polarisation-controller states of each launch group (see Data availability statement).

The FMF was coupled into the multimode fibre (MMF; Thorlabs FG025LJA) via a free-space gap. The two fibre ends were mounted facing each other in a Fujikura 70S fusion splicer operated in manual-alignment mode, with no fusion arc applied. Its positioning stages and built-in microscope thus served as a three-axis stage, used to set and measure the transverse core offset ($\Delta x$, $\Delta y$) and the axial gap ($\Delta z$) between the fibres. This controlled offset launch, together with the controller states, determined the modal superposition excited in the MMF. We measured 250 launch groups with random launch position and with transverse offsets $\Delta x$ and $\Delta y$ between $-9$ and $9\,\mu$m and an axial gap $\Delta z$ between 25 and $45\,\mu$m. Within each group the polarisation controller was set to 20 random positions.

The MMF output was collimated by an aspheric lens (L1; Thorlabs C220TME, f=11 mm), and a linear polariser (Thorlabs LPVIS050-MP2) selected one polarisation component. A cube beamsplitter (BS) divided the field between a reconstruction channel and a holographic channel. In the reconstruction channel, the transmitted field illuminated a reflective phase-only spatial light modulator (HOLOEYE ERIS-1.1) displaying the optimised ROI-constrained mask (seed 12, Supplementary Note~4) at its native sampling ($626\times626$ pixels, 8\,$\mu$m pitch, 256 phase levels). The SLM was 105\,mm from L1 and camera C1 60\,mm from the SLM. The mask was centred on the beam, whose position on the SLM was found beforehand by displaying a circular aperture, illuminating it with a single-spatial-mode beam from the laser and moving it until the diffraction pattern was circularly symmetric (Supplementary Note~4). Because the operator is calibrated from holography, reconstruction does not rely on this design model. The returned field was reflected by the BS and propagated in free space to camera C1 (Ximea MQ013RG-E2), which recorded the symmetry-broken-plane intensity.

In the holographic channel, the field reflected by the BS was imaged onto camera C2 (Thorlabs LP126MU) by a second lens (L2; Thorlabs AC254-500-A, f=500 mm); L1 and L2 form a 4-f imaging system that projects the fibre near field onto C2 with a magnification of about 45. Reference-arm light was collimated (Thorlabs F810FC-635) and reached C2 at a non-collinear angle. The holograms were used as a source of "ground truth" about EM field distribution for calibration of the setup and independent testing of the setup. After calibration, reconstruction of each unknown field used only its symmetry-broken-plane intensity.

The diffuser configuration was identical except in the reconstruction channel: the SLM was removed, and the field transmitted by the BS passed through a ground-glass diffuser (Thorlabs DG10-220-MD) directly onto C1, which faced the BS.

\subsection{Experimental calibration and evaluation}

The experimental camera-plane complex modes were calibrated by optimisation using holography-derived LP labels. A fixed linear phase correction, LP synthesis basis, camera-mode operator and regularised Cholesky factor were then used for held-out inference. Both acquisitions share the same 42-component LP synthesis basis, computed for the FG025LJA fibre at the source wavelength of 632.8 nm; for each acquisition, the shared phase correction, reference-quality thresholds and camera-plane modal fields were fitted on calibration references only. The release supplies the historical inputs, scripts and frozen artefacts for replaying inference and the experimental selection and calibration workflow. Recalibration can start from the three included warm-start checkpoints; their earlier training lineage is not included (Data and Code availability).

Each symmetry-broken-plane image is dark-corrected, normalised and pooled to $256\times256$ pixels, and each reconstructed field is synthesised on a $64\times64$ grid. For each image, let $P$ be the resulting $65,536\times1,764$ PMP matrix and $G=P^{\mathrm H}P$. The coefficient products are estimated as
\begin{equation}
\widehat q=(G+\lambda_{\mathrm r}I)^{-1}P^{\mathrm H}i_d,\qquad
\lambda_{\mathrm r}=10^{-3}\operatorname{tr}(G)/1764.
\label{eq:experimental-inverse}
\end{equation}
After reshaping, the estimated modal-product matrix is Hermitian by construction: the PMP columns occur in conjugate pairs, the intensity data are real, and the ridge penalty preserves this structure. Its leading eigenpair gives the rank-one positive-semidefinite estimate and the modal coefficients, which synthesise the near field. No Hermitian symmetrisation is required in exact arithmetic; the supplied code averages the matrix with its adjoint only as a safeguard against floating-point round-off. The rectangular PMP measurement matrix $P$ is distinct from this Hermitian product matrix; its Gram matrix $G=P^{\mathrm H}P$ is also Hermitian by construction. Only symmetry-broken-plane intensity enters this inference.

For flattened reconstructed and corrected holographic fields $u$ and $v$, respectively, we use
\begin{equation}
\alpha=\frac{u^{\mathrm H}v}{u^{\mathrm H}u},\qquad
\mathrm{NMSE}=\frac{\|\alpha u-v\|_2^2}{\|v\|_2^2},\qquad
\gamma=\frac{|u^{\mathrm H}v|}{\|u\|_2\|v\|_2}.
\label{eq:experimental-metrics}
\end{equation}
The alignment removes global amplitude scale and phase while absolute optical power and phase are not established. Per sample, $\mathrm{NMSE}=1-\gamma^2$. Because $u$ lies in the span of the 42 LP synthesis fields, replacing $v$ by its orthogonal projection $v_{\mathrm{LP}}$ onto that span gives the error relative to the 42-mode content of the reference, $\mathrm{NMSE}_{\mathrm{LP}}=1-(1-\mathrm{NMSE})/\rho^2$, where $\rho=\|v_{\mathrm{LP}}\|_2/\|v\|_2$ is the LP coherence of the reference and $1-\rho^2$ its out-of-subspace fraction. The reported medians use the lower-middle observation for even counts. All reported scores use these fixed eligibility masks while the excluded references are characterised in Supplementary Note~6.  This Note~6 also gives preprocessing, exclusion counts, coefficient metrics, spectral cutoff and reproducibility details.

\subsection{Inference speed}

Inference speed was measured with \texttt{reporting/benchmarks/benchmark\_inference\_cuda.py} on a workstation with an Intel Core i7-8700 CPU and an NVIDIA GeForce RTX 3090 (24 GB; driver 581.29, CUDA runtime 12.8, PyTorch 2.11, Python 3.12, Windows 11) with the two reported 632.8 nm operators. For each operator, the inverse is formed once and stored in single precision as $M^2=1,764$ real rows (0.43 GiB). Per frame, a raw 16-bit $1,024\times1,280$ frame is uploaded from pinned host memory, dark-subtracted, cropped, normalised and pooled to $256\times256$ as in the calibration, multiplied by the stored inverse, assembled into the $42\times42$ product matrix and reduced to its leading eigenpair, and the weights are returned to host memory. For this matrix size the eigendecomposition is faster on the CPU than on the GPU and was performed there. Batch-1 latencies are medians of 300 repetitions after 30 warm-up runs. On all 500 held-out images of each acquisition, the single-precision GPU reconstructions agree with the CPU reference implementation to NMSE $\leq3.7\times10^{-9}$ and reproduce the eligible medians of 4.61\% and 9.09\%.

\subsection{Mask optimisation and matched controls}

Free-space propagation, the thin lens and phase modulation are implemented as differentiable complex-valued operations (Supplementary Note~3). The source wavelength is 632.8 nm throughout; the numerical simulations use its rounded value of 633 nm. The simulations use 42 LP modes with core radius $12.5\,\mu$m and numerical aperture 0.10, a $1024\times1024$ phase plane at the $8\,\mu$m SLM pitch with 256 phase levels, and a $1024\times1024$ camera at $5.3\,\mu$m pitch, 68.59 mm behind the mask. The mask is parameterised by an unwrapped $128\times128$ control grid, bilinearly interpolated onto the phase plane and quantised to 256 levels in every forward pass with a straight-through derivative. The ROI is an order-four super-Gaussian with half weight at radius 2 mm; captured fractions use pixel-area quadrature and the power incident on the mask. Each optimisation starts from a weak random phase (RMS $\approx0.04$ rad), optimises a regularised, trace-normalised log-determinant of the PMP Gram matrix for 100 epochs and then minimises $\log\kappa$ for 300 epochs, evaluated on 8,192 ROI pixels resampled at every epoch. The mean-capture constraint is enforced with an augmented Lagrangian, and both constraints with projected steps and backtracking. The checkpoint with the lowest training condition number that satisfies the constraints is kept; final condition numbers are recomputed on all camera pixels.

The matched controls share the optical system, phase quantisation and incident-power normalisation: a flat-phase SLM, 144 tiled random masks (24 at each cell size of 1, 2, 4, 8, 16 and 32 pixels) and 72 smooth random masks on the same $128\times128$ control grid. The peak-to-mean intensity of the worst of the 42 basis modes was evaluated after optimisation and was not used for training or checkpoint selection. Supplementary Note~4 gives the ensembles and numerical checks.

\subsection{Field recovery and evaluation}

The experimental numerical-rank convention is given in Supplementary Note~6; the simulations use full-rank QR as described below. In this work, we recover modal coefficients by rank-one positive-semidefinite projection of the estimated product matrix, using all pairwise products for greater noise robustness. Direct computation from its diagonal and a reference row is also feasible and faster, but its phase estimates become sensitive to noise when the reference coefficient is weak. Supplementary Note~2 gives the two coefficient-extraction formulas. Analogue-to-digital units (ADU) denote digitised camera counts. Modal coefficients are recovered up to a global phase and the field is synthesised in the original basis.

\subsection{Simulated reconstruction and noise}

The simulations of Fig.~\ref{fig:direct-statistics} use ten core radii (4.1, 6, 7, 8, 9, 10, 11, 12, 12.5 and 14\,$\mu$m), giving 6, 10, 12, 17, 23, 25, 30, 38, 42 and 51 guided scalar LP modes, and fifteen SNR targets ($-15$, $-10$, $-5$, 0, 2, 5, 8, 11, 13, 17, 20, 23, 26, 29 and 32 dB). Each radius has 1,000 circular-complex-Gaussian coefficient vectors normalised to unit norm, paired across SNR levels with independent pixel noise at each level, together with noiseless controls. Changing the core radius changes the physical basis as well as the mode count. For each radius, the complete unregularised inverse $\mathbf{\Upsilon}^{+}$ is formed with full-rank column-pivoted QR using all $1024^2$ camera pixels and all $M^2$ ordered products, and reused at every noise level; this is the zero-ridge case of Eq.~\ref{eq:linear-reconstruction}. For the five targets below 5 dB and the radii with 23 or more modes, the same least-squares estimate was computed by a QR factorisation of the full PMP matrix on a GPU, without storing the inverse. Each image requires one inverse--intensity multiplication, Hermitian symmetrisation and rank-one projection, with a negative leading eigenvalue clipped to zero.

The detector has a 10-bit analogue-to-digital converter, a 128 ADU pedestal and a 700 ADU peak signal per field. Independent Gaussian read noise is calibrated per field to the realised SNR after rounding to integer ADU, and negative pedestal-subtracted values are retained. For the ten targets from 5 to 32 dB the values are also clipped to the 10-bit range, which affected at most $1.1\times10^{-5}$ of the pixels of any image; for the five lower targets clipping is not applied, so that the noise remains Gaussian even where its standard deviation exceeds the pedestal. This is an equal-SNR comparison with per-field exposure, not a fixed-exposure or photon-limited prediction. For clean camera intensity $I_p$ and noisy intensity $\widetilde I_p$, the realised SNR is
\begin{equation}
\mathrm{SNR}=10\log_{10}\frac{\sum_p I_p^2}{\sum_p(\widetilde I_p-I_p)^2},
\end{equation}
with both sums over every sensor pixel after quantisation and pedestal subtraction. This full-sensor energy ratio is lower than the peak-based definition $20\log_{10}(\max_p I_p/\sigma)$, the brightest pixel over the standard deviation $\sigma$ of the added Gaussian noise, because the light occupies only part of the sensor and the speckle has a high peak-to-mean ratio. For the simulated fields the two differ by a nearly constant offset, $20\log_{10}(\max_p I_p/\sigma)\approx\mathrm{SNR}+24.9$\,dB (median over all 150,000 noisy trials; 24.7--25.3\,dB for the individual mode counts), so that SNRs of $-15$, 5 and 32\,dB correspond to $\max_p I_p/\sigma\approx3$, 31 and 700. Fig.~\ref{fig:direct-statistics} and Supplementary Note~5 give both scales. Coefficient and field errors are
\begin{equation}
\epsilon_c=\min_{|a|=1}\frac{\|a\widehat{\mathbf C}-\mathbf C\|_2^2}{\|\mathbf C\|_2^2},\qquad
\epsilon_E=\min_{|b|=1}\frac{\|b\widehat E-E\|_2^2}{\|E\|_2^2},
\end{equation}
with the field norm evaluated in the sampled fibre plane. Each metric removes only its own unobservable global phase and fits no amplitude; relative $L^2$ errors are their square roots. These definitions differ from the complex-scalar-aligned experimental metrics. Displayed examples are the trials whose coefficient error is closest to the ensemble median. The full-sensor simulation latency on an Apple Metal device is reported in Supplementary Note~5.

\subsection{Figure creation}
Figure 1 was assembled from data generated using the simulation framework developed for this study, including near-field intensity and complex-field distributions, phase masks, symmetry-broken intensity patterns, and modal weights of distinct input fields and their reconstructions. The figure layout and composition were assisted by \textit{Automating Schematic Figure Making} \cite{zhao2026automating}. Figure 5 was initially designed and drawn by the author in Inkscape, with subsequent colour styling assisted by the same approach \cite{zhao2026automating}.

\section{Data availability}
The processed data, frozen operators, historical calibration inputs, saved predictions and validation records are available from Zenodo (\url{https://doi.org/10.5281/zenodo.23785122}).

\section{Code availability}
The same Zenodo record (\url{https://doi.org/10.5281/zenodo.23785122}) contains the code and inputs for reproducing frozen-operator inference, experimental selection and calibration, and generating the paper figures.

\section{Acknowledgements}
E.M. acknowledges support from the Royal Society (IF\textbackslash{}R2\textbackslash{}25 2078) and the UK Multidisciplinary Centre for Neuromorphic Computing (UKRI982).

\section{Author contributions}
E.M. conceived the idea, developed the theory and methods, designed and performed the simulations and experiments, analysed the data, wrote the software and wrote the manuscript.

\section{Competing interests}
The author declares no competing interests.

\bibliographystyle{ieeetr}
\bibliography{references}

@article{manuylovich2020fast,
  title={Fast mode decomposition in few-mode fibers},
  author={Manuylovich, Egor S and Dvoyrin, Vladislav V and Turitsyn, Sergei K},
  journal={Nature communications},
  volume={11},
  number={1},
  pages={5507},
  year={2020},
  publisher={Nature Publishing Group UK London},
  doi={10.1038/s41467-020-19323-6}
}

@article{manuylovich2021intensity,
  title={Intensity-only-measurement mode decomposition in few-mode fibers},
  author={Manuylovich, Egor and Donodin, Aleksandr and Turitsyn, Sergei},
  journal={Optics Express},
  volume={29},
  number={22},
  pages={36769--36783},
  year={2021},
  publisher={Optica Publishing Group},
  doi={10.1364/OE.437907}
}

@article{norris2020all,
  title={An all-photonic focal-plane wavefront sensor},
  author={Norris, Barnaby RM and Wei, Jin and Betters, Christopher H and Wong, Alison and Leon-Saval, Sergio G},
  journal={Nature Communications},
  volume={11},
  number={1},
  pages={5335},
  year={2020},
  publisher={Nature Publishing Group UK London},
  doi={10.1038/s41467-020-19117-w}
}

@article{cao2024high,
  title={High-resolution, large field-of-view label-free imaging via aberration-corrected, closed-form complex field reconstruction},
  author={Cao, Ruizhi and Shen, Cheng and Yang, Changhuei},
  journal={Nature Communications},
  volume={15},
  number={1},
  pages={4713},
  year={2024},
  publisher={Nature Publishing Group UK London},
  doi={10.1038/s41467-024-49126-y}
}

@article{grohs2020phase,
  title={Phase retrieval: uniqueness and stability},
  author={Grohs, Philipp and Koppensteiner, Sarah and Rathmair, Martin},
  journal={SIAM Review},
  volume={62},
  number={2},
  pages={301--350},
  year={2020},
  publisher={SIAM}
}

@article{shechtman2015phase,
  title={Phase retrieval with application to optical imaging: a contemporary overview},
  author={Shechtman, Yoav and Eldar, Yonina C and Cohen, Oren and Chapman, Henry Nicholas and Miao, Jianwei and Segev, Mordechai},
  journal={IEEE signal processing magazine},
  volume={32},
  number={3},
  pages={87--109},
  year={2015},
  publisher={IEEE}
}

@article{fienup1982phase,
  author = {Fienup, James R.},
  title = {Phase retrieval algorithms: a comparison},
  journal = {Applied Optics},
  volume = {21},
  number = {15},
  pages = {2758--2769},
  year = {1982},
  doi = {10.1364/AO.21.002758}
}

@article{gonsalves1982diversity,
  author = {Gonsalves, Robert A.},
  title = {Phase retrieval and diversity in adaptive optics},
  journal = {Optical Engineering},
  volume = {21},
  number = {5},
  pages = {215829},
  year = {1982},
  doi = {10.1117/12.7972989}
}

@article{candes2015coded,
  author = {Cand{\`e}s, Emmanuel J. and Li, Xiaodong and Soltanolkotabi, Mahdi},
  title = {Phase retrieval from coded diffraction patterns},
  journal = {Applied and Computational Harmonic Analysis},
  volume = {39},
  number = {2},
  pages = {277--299},
  year = {2015},
  doi = {10.1016/j.acha.2014.09.004}
}

@article{martinache2013asymmetric,
  author = {Martinache, Frantz},
  title = {The asymmetric pupil {Fourier} wavefront sensor},
  journal = {Publications of the Astronomical Society of the Pacific},
  volume = {125},
  pages = {422--430},
  year = {2013},
  doi = {10.1086/670670}
}

@article{gao2026single,
  title   = {Single-shot, reference-less computational wavefront sensing for complex optical fields},
  author  = {Gao, Yunhui and Cao, Liangcai and Tsai, Din Ping},
  journal = {Light: Science \& Applications},
  volume  = {15},
  pages   = {174},
  year    = {2026},
  doi     = {10.1038/s41377-026-02241-5}
}

@article{stutzki2011high,
  title = {High-speed modal decomposition of mode instabilities in high-power fiber lasers},
  author = {Stutzki, Fabian and Otto, Hans-J{\"u}rgen and Jansen, Florian and Gaida, Christian and Jauregui, Cesar and Limpert, Jens and T{\"u}nnermann, Andreas},
  journal = {Optics Letters},
  volume = {36},
  number = {23},
  pages = {4572--4574},
  year = {2011},
  doi = {10.1364/OL.36.004572}
}

@article{flamm2013digital,
  title = {All-Digital Holographic Tool for Mode Excitation and Analysis in Optical Fibers},
  author = {Flamm, Daniel and Schulze, Christian and Naidoo, Darryl and Schr{\"o}ter, Siegmund and Forbes, Andrew and Duparr{\'e}, Michael},
  journal = {Journal of Lightwave Technology},
  volume = {31},
  number = {7},
  pages = {1023--1032},
  year = {2013},
  doi = {10.1109/JLT.2013.2240258}
}

@article{nicholson2008spatially,
  title = {Spatially and spectrally resolved imaging of modal content in large-mode-area fibers},
  author = {Nicholson, J. W. and Yablon, A. D. and Ramachandran, S. and Ghalmi, S.},
  journal = {Optics Express},
  volume = {16},
  number = {10},
  pages = {7233--7243},
  year = {2008},
  doi = {10.1364/OE.16.007233}
}

@article{nguyen2012modal,
  title = {Modal decomposition technique for multimode fibers},
  author = {Nguyen, Duc Minh and Blin, St{\'e}phane and Nguyen, Thanh Nam and Le, Sy Dat and Provino, Laurent and Thual, Monique and Chartier, Thierry},
  journal = {Applied Optics},
  volume = {51},
  number = {4},
  pages = {450--456},
  year = {2012},
  doi = {10.1364/AO.51.000450}
}

@article{schimpf2011cross,
  title = {Cross-correlated ({C}$^2$) imaging of fiber and waveguide modes},
  author = {Schimpf, D. N. and Barankov, R. A. and Ramachandran, S.},
  journal = {Optics Express},
  volume = {19},
  number = {14},
  pages = {13008--13019},
  year = {2011},
  doi = {10.1364/OE.19.013008}
}

@article{demas2014subsecond,
  title = {Sub-second mode measurement of fibers using {C}$^2$ imaging},
  author = {Demas, Jeffrey and Ramachandran, Siddharth},
  journal = {Optics Express},
  volume = {22},
  number = {19},
  pages = {23043--23056},
  year = {2014},
  doi = {10.1364/OE.22.023043}
}

@article{kaiser2009complete,
  title = {Complete modal decomposition for optical fibers using {CGH}-based correlation filters},
  author = {Kaiser, Thomas and Flamm, Daniel and Schr{\"o}ter, Siegmund and Duparr{\'e}, Michael},
  journal = {Optics Express},
  volume = {17},
  number = {11},
  pages = {9347--9356},
  year = {2009},
  doi = {10.1364/OE.17.009347}
}

@article{flamm2012mode,
  title = {Mode analysis with a spatial light modulator as a correlation filter},
  author = {Flamm, Daniel and Naidoo, Darryl and Schulze, Christian and Forbes, Andrew and Duparr{\'e}, Michael},
  journal = {Optics Letters},
  volume = {37},
  number = {13},
  pages = {2478--2480},
  year = {2012},
  doi = {10.1364/OL.37.002478}
}

@article{gervaziev2021mode,
  title = {Mode decomposition of multimode optical fiber beams by phase-only spatial light modulator},
  author = {Gervaziev, M. D. and Zhdanov, I. and Kharenko, D. S. and Gonta, V. A. and Volosi, V. M. and Podivilov, E. V. and Babin, S. A. and Wabnitz, S.},
  journal = {Laser Physics Letters},
  volume = {18},
  number = {1},
  pages = {015101},
  year = {2021},
  doi = {10.1088/1612-202X/abcf27}
}

@article{lu2013fast,
  title = {Fast and accurate modal decomposition of multimode fiber based on stochastic parallel gradient descent algorithm},
  author = {L{\"u}, Haibin and Zhou, Pu and Wang, Xiaolin and Jiang, Zongfu},
  journal = {Applied Optics},
  volume = {52},
  number = {12},
  pages = {2905--2908},
  year = {2013},
  doi = {10.1364/AO.52.002905}
}

@article{li2017hybrid,
  title = {Multimode fiber modal decomposition based on hybrid genetic global optimization algorithm},
  author = {Li, Lei and Leng, Jinyong and Zhou, Pu and Chen, Jinbao},
  journal = {Optics Express},
  volume = {25},
  number = {17},
  pages = {19680--19690},
  year = {2017},
  doi = {10.1364/OE.25.019680}
}

@article{an2019learning,
  title = {Learning to decompose the modes in few-mode fibers with deep convolutional neural network},
  author = {An, Yi and Huang, Liangjin and Li, Jun and Leng, Jinyong and Yang, Lijia and Zhou, Pu},
  journal = {Optics Express},
  volume = {27},
  number = {7},
  pages = {10127--10137},
  year = {2019},
  doi = {10.1364/OE.27.010127}
}

@article{kim2023convolutional,
  title = {Convolutional Neural Network Combined With Stochastic Parallel Gradient Descent to Decompose Fiber Modes Based on Far-Field Measurements},
  author = {Kim, Byungho and Na, Jeongkyun and Jeong, Yoonchan},
  journal = {Journal of Lightwave Technology},
  volume = {41},
  number = {18},
  pages = {5973--5982},
  year = {2023},
  doi = {10.1109/JLT.2023.3276366}
}

@article{yan2019modal,
  title = {Modal decomposition for few mode fibers using the fractional {Fourier} system},
  author = {Yan, Wei and Xu, Xiaojun and Wang, Jianguo},
  journal = {Optics Express},
  volume = {27},
  number = {10},
  pages = {13871--13883},
  year = {2019},
  doi = {10.1364/OE.27.013871}
}

@article{manuylovich2025sparse,
  title = {Sparse intensity sampling for ultrafast full-field reconstruction in low-dimensional photonic systems},
  author = {Manuylovich, Egor},
  journal = {Communications Physics},
  volume = {8},
  number = {1},
  pages = {149},
  year = {2025},
  doi = {10.1038/s42005-025-02079-0}
}

@article{zou2025reference,
  title = {Reference-free mode decomposition of few-mode fibers via coherent modulation imaging},
  author = {Zou, Chunyu and Veetil, Suhas P. and Rong, Na and Jiang, Zhilong and Liu, Cheng and Zhu, Jianqiang},
  journal = {Optics Letters},
  volume = {50},
  number = {11},
  pages = {3648--3651},
  year = {2025},
  doi = {10.1364/OL.553224}
}

@article{rothe2020deep,
  title = {Deep Learning for Computational Mode Decomposition in Optical Fibers},
  author = {Rothe, Stefan and Zhang, Qian and Koukourakis, Nektarios and Czarske, J{\"u}rgen W.},
  journal = {Applied Sciences},
  volume = {10},
  number = {4},
  pages = {1367},
  year = {2020},
  doi = {10.3390/app10041367}
}

@article{badon2016smart,
  title = {Smart optical coherence tomography for ultra-deep imaging through highly scattering media},
  author = {Badon, Amaury and Li, Dayan and Lerosey, Geoffroy and Boccara, A. Claude and Fink, Mathias and Aubry, Alexandre},
  journal = {Science Advances},
  volume = {2},
  number = {11},
  pages = {e1600370},
  year = {2016},
  doi = {10.1126/sciadv.1600370}
}

@article{wen2023single,
  title = {Single multimode fibre for in vivo light-field-encoded endoscopic imaging},
  author = {Wen, Zhong and Dong, Zhenyu and Deng, Qilin and Pang, Chenlei and Kaminski, Clemens F. and Xu, Xiaorong and Yan, Huihui and Wang, Liqiang and Liu, Songguo and Tang, Jianbin and Chen, Wei and Liu, Xu and Yang, Qing},
  journal = {Nature Photonics},
  volume = {17},
  pages = {679--687},
  year = {2023},
  doi = {10.1038/s41566-023-01240-x}
}

@article{zuo2022deep,
  title = {Deep learning in optical metrology: a review},
  author = {Zuo, Chao and Qian, Jiaming and Feng, Shijie and Yin, Wei and Li, Yixuan and Fan, Pengfei and Han, Jing and Qian, Kemao and Chen, Qian},
  journal = {Light: Science \& Applications},
  volume = {11},
  pages = {39},
  year = {2022},
  doi = {10.1038/s41377-022-00714-x}
}

@article{zhao2026automating,
  title   = {Automating schematic figure making},
  author  = {Zhao, Jinchen and Wright, Logan},
  journal = {lgwrightlab.github.io},
  year    = {2026},
  month   = {Oct},
  url     = {https://lgwrightlab.github.io/schematic-figure/}
}

\clearpage
\makeatletter
\def\@ptsize{1}
\input{size11.clo}
\makeatother
\normalsize
\newgeometry{margin=2.3cm}
\setcounter{section}{0}
\setcounter{subsection}{0}
\setcounter{equation}{0}
\setcounter{figure}{0}
\setcounter{table}{0}
\setcounter{secnumdepth}{3}
\let\addcontentsline\arxivAddContentsLine

\begin{center}
  {\LARGE\bfseries Supplementary Information\par}
  \vspace{0.6em}
  {\large \arxivArticleTitle\par}
  \vspace{0.6em}
  \articleAuthor
\end{center}
\renewcommand{\figurename}{Supplementary Fig.}
\renewcommand{\tablename}{Supplementary Table}

\tableofcontents

\section{Supplementary Note 1: Complete derivation of the PMP matrix}

\subsection{Field and intensity in a constrained modal basis}

Consider a single-polarisation optical field constrained to $M$ known modes $\Psi_j$. At the fibre facet,
\begin{equation}
E(x,y,z_{\mathrm{fib}})=\sum_{j=1}^{M}C_j\Psi_j(x,y,z_{\mathrm{fib}}),
\label{eq:S-field}
\end{equation}
where the harmonic factor $\exp(i\omega t)$ is omitted. A linear optical operator $\Lop$ maps the field from the fibre facet to the camera. It may contain free-space propagation, lenses, phase elements and other linear components. Thus,
\begin{equation}
\Psi_j^{\mathrm{cam}}(x,y;\Lop)=\Lop[\Psi_j(x,y,z_{\mathrm{fib}})].
\label{eq:S-camera-mode}
\end{equation}
The camera records
\begin{align}
I(x,y,z_{\mathrm{cam}})
&=\left|\sum_{j=1}^{M}C_j\Psi_j^{\mathrm{cam}}(x,y;\Lop)\right|^2 \\
&=\sum_{j=1}^{M}\sum_{k=1}^{M}C_j\overline{C_k}
\Psi_j^{\mathrm{cam}}(x,y;\Lop)
\overline{\Psi_k^{\mathrm{cam}}(x,y;\Lop)}.
\label{eq:S-intensity}
\end{align}

After sampling $N=N_xN_y$ camera pixels and flattening the images, define
\begin{equation}
s_{M(j-1)+k}=C_j\overline{C_k},\qquad
\boldsymbol{\upsilon}_{M(j-1)+k}(\Lop)=
\boldsymbol{\psi}^{\mathrm{cam}}_j(\Lop)\circ
\overline{\boldsymbol{\psi}^{\mathrm{cam}}_k(\Lop)},
\label{eq:S-change-variables}
\end{equation}
where $\circ$ is the Hadamard product. Collecting the $M^2$ pairwise mode products as columns gives
\begin{equation}
\mathbf{\Upsilon}(\Lop)=
\begin{bmatrix}
|&|&&|\\
\boldsymbol{\upsilon}_1&\boldsymbol{\upsilon}_2&\cdots&\boldsymbol{\upsilon}_{M^2}\\
|&|&&|
\end{bmatrix},
\qquad
\mathbf{i}=\mathbf{\Upsilon}(\Lop)\mathbf{s}.
\label{eq:S-PMP}
\end{equation}
The ordering used here is
\begin{equation}
\mathbf{s}=\left[C_1\overline{C_1},C_1\overline{C_2},\ldots,
C_1\overline{C_M},C_2\overline{C_1},\ldots,C_M\overline{C_M}\right]^{\mathsf T}.
\label{eq:S-vector}
\end{equation}

\subsection{Degeneracy and conditioning}

For the ten-mode example in main-text Fig.~2a, 45 of the 100 near-field singular values vanish because a product of two real mode profiles is unchanged when the modes are exchanged; seven further singular values lie below $10^{-4}$ of the largest. The far-field matrix has $\kappa\approx3\times10^{3}$, so propagation alone changes observability before the engineered mask is introduced.

If two PMP columns coincide, detector intensity cannot distinguish their contributions to Eq.~\ref{eq:S-PMP}. If the columns are nearly linearly dependent, the distinction is unstable under noise. We therefore use
\begin{equation}
\mathcal{D}(\Lop)=\kappa[\mathbf{\Upsilon}(\Lop)]
=\frac{\sigma_{\max}[\mathbf{\Upsilon}(\Lop)]}
{\sigma_{\min}[\mathbf{\Upsilon}(\Lop)]}
\label{eq:S-condition}
\end{equation}
as a discrete measure of measurement degeneracy. Condition numbers refer to the complex PMP matrix with all ordered products. Input modes are individually normalised to unit discretised energy (Supplementary Note~4); the experimental rank cutoff is specified in Supplementary Note~6.

A continuous analogue is the Gram matrix of the mode-product functions. For a family of PMP functions $u_p(x,y)$, that form is
\begin{equation}
G_{pq}=\iint u_p(x,y)\overline{u_q(x,y)}\,dx\,dy,
\qquad \kappa(\mathbf G)=\kappa(\mathbf U)^2
\label{eq:S-Gram}
\end{equation}
for a full-rank discretisation $\mathbf U$ with consistent quadrature.

\subsection{Near-field degeneracies of the 42-mode fibre}

To illustrate which pairwise products a near-field intensity measurement cannot separate, we computed the 42 guided LP modes of the simulated fibre (core radius 12.5\,$\mu$m, numerical aperture 0.10, cladding index 1.4563, wavelength 632.8 nm (rounded to 633 nm in the numerical model)) with the LP solver used throughout this work. The solver returns real fields: $\mathrm{LP}_{0m}$ modes and, for $l>0$, the partners $\mathrm{LP}_{lm}^{a}\propto\cos l\phi$ and $\mathrm{LP}_{lm}^{b}\propto\sin l\phi$. We formed all $42^2=1,764$ ordered products $\psi_k\overline{\psi_j}$, the columns of the PMP matrix (Eq.~\ref{eq:S-PMP}), and, for comparison, the unconjugated products $\psi_k\psi_j$.

\emph{Origin in the cylindrical symmetry of the waveguide.} Every LP mode separates into a radial profile and an angular factor, $\psi=R_{lm}(r)\,\Theta_l(\phi)$ with $\Theta_l\in\{1,\cos l\phi,\sin l\phi\}$, so every product separates into a radial part $R_{lm}R_{l'm'}$ and an angular part $\Theta_l\Theta_{l'}$. The angular parts multiply like Fourier harmonics: modes of azimuthal orders $l$ and $l'$ give harmonics of orders $l+l'$ and $|l-l'|$ only, e.g.\ $\cos l\phi\,\sin l'\phi=\tfrac12[\sin(l+l')\phi-\sin(l-l')\phi]$. Because the waveguide is rotationally symmetric, many different mode pairs therefore produce exactly the same angular pattern: products of two lower-order angular components reproduce a single higher-order one, so that, for instance, $\mathrm{LP}_{1m}$ with $\mathrm{LP}_{2m'}$ yields third-order (and first-order) harmonics for any radial indices $m$, $m'$, just like a product involving a third-order mode. Two products are then distinguished only by their radial parts.

(i) When the radial parts coincide as well, the products are \emph{exactly} identical. This happens whenever the angular partners of two modes of the same order are exchanged: $\mathrm{LP}_{lm}^{a}\overline{\mathrm{LP}_{lm'}^{b}}$ and $\mathrm{LP}_{lm}^{b}\overline{\mathrm{LP}_{lm'}^{a}}$ both equal $\tfrac12R_{lm}R_{lm'}\sin 2l\phi$. These exact degeneracies are imposed by the cylindrical symmetry alone, and no reconstruction algorithm can separate them from a near-field image; they are precisely what a symmetry-breaking measurement removes.

(ii) When only the angular parts coincide, the products differ through radial profiles that, for a step-index fibre, are similar Bessel envelopes; they are then close but not identical. Examples are $\mathrm{LP}_{01}$ and $\mathrm{LP}_{02}$ multiplied by the same $\mathrm{LP}_{81}^{b}$ mode, which share the eighth-order pattern $\sin 8\phi$ of $\mathrm{LP}_{41}^{a}\overline{\mathrm{LP}_{41}^{b}}$, and $\mathrm{LP}_{41}\overline{\mathrm{LP}_{52}}$ and $\mathrm{LP}_{42}\overline{\mathrm{LP}_{51}}$, which both contain the harmonics of orders 1 and 9. These near-degeneracies are also rooted in the rotational symmetry; the residual difference comes only from the radial components.

\emph{Trivial duplicates.} Because the fields are real, $\psi_k\psi_j=\psi_k\overline{\psi_j}$, so the unconjugated products add nothing, and $\psi_k\overline{\psi_j}=\psi_j\overline{\psi_k}$: every product of two different modes appears as two identical ordered members, $(k,j)$ and $(j,k)$. These 861 trivial pairs, and the 42 single diagonal products $|\psi_k|^2$, reflect the Hermitian structure of the coefficient-product matrix, which the reconstruction already uses, and are omitted from the catalogue of non-trivial groups; they nevertheless contribute null directions to the full near-field PMP matrix. Both identities hold to machine precision. The analysis therefore uses the 903 distinct products, one per unordered mode pair, and a group is counted as non-trivial only if it contains at least two different mode pairs.

\emph{Similarity criterion.} For every pair of distinct products $u_p$, $u_q$ on a $32\times32$ near-field grid, the similarity is $|\langle u_p,u_q\rangle|/(\|u_p\|\,\|u_q\|)$. A group is a maximal set in which every pair has similarity above 0.95, i.e.\ a maximal clique of the graph that links products above this threshold, not merely a connected component. All groups were recomputed from $256\times256$ solver fields, without upsampling, and all pass the threshold at both resolutions.

\begin{figure}[!ht]
\centering
\includegraphics[width=\linewidth]{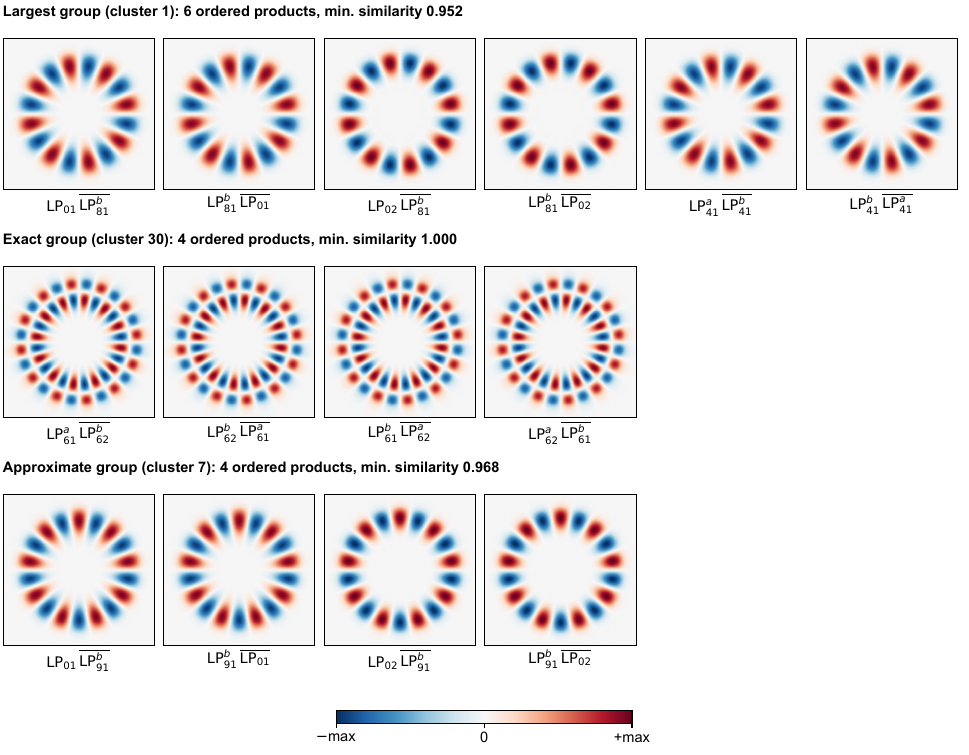}
\caption{Near-field degeneracies of pairwise LP-mode products in the 42-mode fibre. Each row shows every ordered member $\psi_k\overline{\psi_j}$ of one group, in which all pairs of distinct products have absolute normalised similarity above 0.95: (top) the only six-member group; (middle) an exact four-member group, $\mathrm{LP}_{61}$ and $\mathrm{LP}_{62}$ with exchanged angular partners; (bottom) a close but not exact four-member group, $\mathrm{LP}_{01}$ and $\mathrm{LP}_{02}$ times $\mathrm{LP}_{91}^{b}$. Index-swapped twins $(k,j)$ and $(j,k)$ are identical because the LP fields are real. $a$, $b$: $\cos l\phi$ and $\sin l\phi$ partners. Products computed on a $256\times256$ grid by the LP solver, cropped to $\pm15\,\mu$m, normalised to unit energy and sign-aligned within each row; one symmetric colour scale. }
\label{fig:S-degenerate}
\end{figure}

\emph{Groups.} There are 30 non-trivial groups, involving 61 of the 903 distinct products. Counted in ordered products $\psi_k\overline{\psi_j}$, one group has six members and 29 have four (Supplementary Fig.~\ref{fig:S-degenerate}). The six-member group consists of $\mathrm{LP}_{01}\,\overline{\mathrm{LP}_{81}^{b}}$, $\mathrm{LP}_{02}\,\overline{\mathrm{LP}_{81}^{b}}$ and $\mathrm{LP}_{41}^{a}\,\overline{\mathrm{LP}_{41}^{b}}$ with their reversed-index copies (smallest pairwise similarity 0.952). The 29 four-member groups, each two mode pairs with their twins, are listed in Supplementary Table~\ref{tab:S-degenerate-groups}. In 15 of them the two products are identical (similarity 1 to within $10^{-6}$): these are the exact symmetry degeneracies (i), in which the angular partners of two modes of the same azimuthal order are exchanged. In 3 more, $\mathrm{LP}_{01}\overline{\mathrm{LP}_{2l,1}^{b}}$ and $\mathrm{LP}_{l1}^{a}\overline{\mathrm{LP}_{l1}^{b}}$ ($l=1,2,3$) share the angular pattern $\sin 2l\phi$, and their radial profiles are almost proportional (similarity above 0.9998). The remaining 11 groups, and the six-member group, are close but not exact (0.951--0.968): they are near-degeneracies of type (ii), with identical angular harmonics and radial profiles that overlap strongly. Exact groups contain indistinguishable product signatures. In approximate groups, the signatures remain distinct but their separation is sensitive to noise (main text, Fig.~2). Supplementary Fig.~\ref{fig:S-degenerate} shows the six-member group, the exact four-member group and the approximate four-member group of highest mode order; the full list with all similarities is in \texttt{data/lp\_product\_degeneracy}.

\begin{table}[!ht]
\centering
\caption{The 29 four-member near-field groups of the 42-mode fibre. Each group contains the two listed products and their reversed-index copies. Similarity: absolute normalised inner product of the two products on the $256\times256$ grid. The six-member group is given in the text.}
\label{tab:S-degenerate-groups}
\small
\begin{tabular}{rccc}
\toprule
Group & Product 1 & Product 2 & Similarity\\
\midrule
2 & $\mathrm{LP}_{01}\,\overline{\mathrm{LP}_{21}^{b}}$ & $\mathrm{LP}_{11}^{a}\,\overline{\mathrm{LP}_{11}^{b}}$ & 1.0000\\
3 & $\mathrm{LP}_{01}\,\overline{\mathrm{LP}_{41}^{b}}$ & $\mathrm{LP}_{21}^{a}\,\overline{\mathrm{LP}_{21}^{b}}$ & 1.0000\\
4 & $\mathrm{LP}_{01}\,\overline{\mathrm{LP}_{61}^{b}}$ & $\mathrm{LP}_{31}^{a}\,\overline{\mathrm{LP}_{31}^{b}}$ & 0.9999\\
5 & $\mathrm{LP}_{01}\,\overline{\mathrm{LP}_{81}^{a}}$ & $\mathrm{LP}_{02}\,\overline{\mathrm{LP}_{81}^{a}}$ & 0.9593\\
6 & $\mathrm{LP}_{01}\,\overline{\mathrm{LP}_{91}^{a}}$ & $\mathrm{LP}_{02}\,\overline{\mathrm{LP}_{91}^{a}}$ & 0.9678\\
7 & $\mathrm{LP}_{01}\,\overline{\mathrm{LP}_{91}^{b}}$ & $\mathrm{LP}_{02}\,\overline{\mathrm{LP}_{91}^{b}}$ & 0.9678\\
8 & $\mathrm{LP}_{11}^{a}\,\overline{\mathrm{LP}_{12}^{b}}$ & $\mathrm{LP}_{11}^{b}\,\overline{\mathrm{LP}_{12}^{a}}$ & 1.0000\\
9 & $\mathrm{LP}_{11}^{a}\,\overline{\mathrm{LP}_{13}^{b}}$ & $\mathrm{LP}_{11}^{b}\,\overline{\mathrm{LP}_{13}^{a}}$ & 1.0000\\
10 & $\mathrm{LP}_{11}^{a}\,\overline{\mathrm{LP}_{14}^{b}}$ & $\mathrm{LP}_{11}^{b}\,\overline{\mathrm{LP}_{14}^{a}}$ & 1.0000\\
11 & $\mathrm{LP}_{12}^{a}\,\overline{\mathrm{LP}_{13}^{b}}$ & $\mathrm{LP}_{12}^{b}\,\overline{\mathrm{LP}_{13}^{a}}$ & 1.0000\\
12 & $\mathrm{LP}_{12}^{a}\,\overline{\mathrm{LP}_{14}^{b}}$ & $\mathrm{LP}_{12}^{b}\,\overline{\mathrm{LP}_{14}^{a}}$ & 1.0000\\
13 & $\mathrm{LP}_{13}^{a}\,\overline{\mathrm{LP}_{14}^{b}}$ & $\mathrm{LP}_{13}^{b}\,\overline{\mathrm{LP}_{14}^{a}}$ & 1.0000\\
14 & $\mathrm{LP}_{21}^{a}\,\overline{\mathrm{LP}_{22}^{b}}$ & $\mathrm{LP}_{21}^{b}\,\overline{\mathrm{LP}_{22}^{a}}$ & 1.0000\\
15 & $\mathrm{LP}_{21}^{a}\,\overline{\mathrm{LP}_{23}^{b}}$ & $\mathrm{LP}_{21}^{b}\,\overline{\mathrm{LP}_{23}^{a}}$ & 1.0000\\
16 & $\mathrm{LP}_{22}^{a}\,\overline{\mathrm{LP}_{23}^{b}}$ & $\mathrm{LP}_{22}^{b}\,\overline{\mathrm{LP}_{23}^{a}}$ & 1.0000\\
17 & $\mathrm{LP}_{31}^{a}\,\overline{\mathrm{LP}_{32}^{b}}$ & $\mathrm{LP}_{31}^{b}\,\overline{\mathrm{LP}_{32}^{a}}$ & 1.0000\\
18 & $\mathrm{LP}_{31}^{a}\,\overline{\mathrm{LP}_{33}^{b}}$ & $\mathrm{LP}_{31}^{b}\,\overline{\mathrm{LP}_{33}^{a}}$ & 1.0000\\
19 & $\mathrm{LP}_{32}^{a}\,\overline{\mathrm{LP}_{33}^{b}}$ & $\mathrm{LP}_{32}^{b}\,\overline{\mathrm{LP}_{33}^{a}}$ & 1.0000\\
20 & $\mathrm{LP}_{41}^{a}\,\overline{\mathrm{LP}_{42}^{b}}$ & $\mathrm{LP}_{41}^{b}\,\overline{\mathrm{LP}_{42}^{a}}$ & 1.0000\\
21 & $\mathrm{LP}_{41}^{a}\,\overline{\mathrm{LP}_{52}^{a}}$ & $\mathrm{LP}_{42}^{a}\,\overline{\mathrm{LP}_{51}^{a}}$ & 0.9512\\
22 & $\mathrm{LP}_{41}^{a}\,\overline{\mathrm{LP}_{52}^{b}}$ & $\mathrm{LP}_{42}^{a}\,\overline{\mathrm{LP}_{51}^{b}}$ & 0.9512\\
23 & $\mathrm{LP}_{41}^{b}\,\overline{\mathrm{LP}_{52}^{a}}$ & $\mathrm{LP}_{42}^{b}\,\overline{\mathrm{LP}_{51}^{a}}$ & 0.9512\\
24 & $\mathrm{LP}_{41}^{b}\,\overline{\mathrm{LP}_{52}^{b}}$ & $\mathrm{LP}_{42}^{b}\,\overline{\mathrm{LP}_{51}^{b}}$ & 0.9512\\
25 & $\mathrm{LP}_{51}^{a}\,\overline{\mathrm{LP}_{52}^{b}}$ & $\mathrm{LP}_{51}^{b}\,\overline{\mathrm{LP}_{52}^{a}}$ & 1.0000\\
26 & $\mathrm{LP}_{51}^{a}\,\overline{\mathrm{LP}_{62}^{a}}$ & $\mathrm{LP}_{52}^{a}\,\overline{\mathrm{LP}_{61}^{a}}$ & 0.9591\\
27 & $\mathrm{LP}_{51}^{a}\,\overline{\mathrm{LP}_{62}^{b}}$ & $\mathrm{LP}_{52}^{a}\,\overline{\mathrm{LP}_{61}^{b}}$ & 0.9591\\
28 & $\mathrm{LP}_{51}^{b}\,\overline{\mathrm{LP}_{62}^{a}}$ & $\mathrm{LP}_{52}^{b}\,\overline{\mathrm{LP}_{61}^{a}}$ & 0.9591\\
29 & $\mathrm{LP}_{51}^{b}\,\overline{\mathrm{LP}_{62}^{b}}$ & $\mathrm{LP}_{52}^{b}\,\overline{\mathrm{LP}_{61}^{b}}$ & 0.9591\\
30 & $\mathrm{LP}_{61}^{a}\,\overline{\mathrm{LP}_{62}^{b}}$ & $\mathrm{LP}_{61}^{b}\,\overline{\mathrm{LP}_{62}^{a}}$ & 1.0000\\
\bottomrule
\end{tabular}
\end{table}

\FloatBarrier

\section{Supplementary Note 2: Detailed reconstruction derivation}

Given the PMP matrix and a measured intensity vector,
\begin{equation}
\widehat{\mathbf s}=\mathbf{\Upsilon}^{\dagger}\mathbf I.
\label{eq:S-pinv}
\end{equation}
one can reshape $\widehat{\mathbf s}$ into $\widehat{\mathbf S}\in\mathbb C^{M\times M}$ so that $\widehat S_{jk}=\widehat s_{M(j-1)+k}$. In the noiseless model,
\begin{equation}
\mathbf S=\mathbf C\mathbf C^{\dagger},
\label{eq:S-rank-one}
\end{equation}
which is Hermitian, positive semidefinite and rank one. The index mapping is
\begin{table}[H]
\centering
\begin{tabular}{c|cccc}
&$\overline C_1$&$\overline C_2$&$\cdots$&$\overline C_M$\\\midrule
$C_1$&$s_1$&$s_2$&$\cdots$&$s_M$\\
$C_2$&$s_{M+1}$&$s_{M+2}$&$\cdots$&$s_{2M}$\\
$\vdots$&$\vdots$&$\vdots$&$\ddots$&$\vdots$\\
$C_M$&$s_{M(M-1)+1}$&$s_{M(M-1)+2}$&$\cdots$&$s_{M^2}$
\end{tabular}
\caption{Mapping between the vectorised PMP coefficients and modal coefficient products.}
\label{tab:S-mapping}
\end{table}

Fixing the unobservable global phase by setting $\arg C_1=0$, the element-wise estimator is
\begin{align}
|\widehat C_j|&=\sqrt{\widehat s_{M(j-1)+j}},\\
\arg(\widehat C_j)&=-\arg(\widehat s_j),\\
\widehat C_j&=\sqrt{\widehat s_{M(j-1)+j}}
\exp[-i\arg(\widehat s_j)].
\label{eq:S-element-recovery}
\end{align}
Finally,
\begin{equation}
\widehat E(x,y,z_{\mathrm{fib}})=\sum_{j=1}^{M}\widehat C_j\Psi_j(x,y,z_{\mathrm{fib}}).
\label{eq:S-field-recovery}
\end{equation}

For real-valued intensity data, conjugate-paired PMP columns and the stated pseudoinverse or scalar-ridge inverse preserve Hermitian symmetry of $\widehat{\mathbf S}$ in exact arithmetic, including in the presence of intensity noise. Noise can nevertheless make this matrix indefinite and depart from rank one. A consistent estimator takes its principal eigenpair to form a rank-one positive-semidefinite approximation:
\begin{equation}
\mathbf H=\widehat{\mathbf S},
\qquad
\widehat{\mathbf C}=\sqrt{\lambda_1}\,\mathbf v_1,
\label{eq:S-eigen-recovery}
\end{equation}
followed by global-phase alignment. This uses all recovered products rather than only the first row and diagonal.

In this work we use rank-one positive-semidefinite projection for its greater noise robustness. Direct element-wise computation of modal weights remains feasible and is faster, because it reads only the diagonal and one reference row instead of computing an eigenpair. Both estimators act on the inverse-PMP estimate of the coefficient products.

The direct MD reference implementation uses $|\widehat C_j|=\sqrt{|\widehat S_{jj}|}$ and $\arg\widehat C_j=-\arg\widehat S_{1j}$, with the first phase set to zero. The absolute value on a noisy diagonal reproduces the reference implementation. When $C_1$ is small, the reference-row products become weak and their phases become noise-sensitive; when $C_1=0$, that row cannot determine the relative phases. Rank-one projection uses all pairwise products and recovers the coefficient vector up to one global phase without requiring a nonzero first coefficient. If that coefficient vanishes, any nonzero recovered coefficient, for example the largest-magnitude one, can define the display gauge.

\section{Supplementary Note 3: Simulation implementation}

\subsection{Optical operator}

The optical system was implemented as a sequence of differentiable free-space
propagation, lens, and phase-modulation operators. For a scalar field
$E(u,v,0)$, free-space propagation over a distance $L$ was modelled using the
Fresnel diffraction integral
\begin{equation}
\mathcal F_L[E](x,y)
=
\frac{\exp(ikL)}{i\lambda L}
\iint
E(u,v,0)
\exp\!\left[
\frac{ik}{2L}
\left((x-u)^2+(y-v)^2\right)
\right]
\,du\,dv,
\label{eq:S-Fresnel}
\end{equation}
where $k=2\pi/\lambda$.

In the numerical implementation, Eq.~\ref{eq:S-Fresnel} is evaluated directly
in separable matrix form rather than through an FFT-based propagator. For one
transverse dimension, the propagation matrix is
\begin{equation}
P_{pm}
=
\Delta x_{\mathrm{in}}
\sqrt{\frac{\exp(ikL)}{i\lambda L}}
\exp\!\left[
\frac{ik}{2L}
\left(x_p^2+u_m^2\right)
\right]
\exp\!\left(
-\frac{ik}{L}x_pu_m
\right),
\label{eq:S-propagation-matrix}
\end{equation}
where $u_m$ and $x_p$ denote the sampled coordinates in the input and output
planes, respectively. The two-dimensional field is then propagated as
\begin{equation}
\mathbf E_{\mathrm{out}}
=
\mathbf P\,
\mathbf E_{\mathrm{in}}\,
\mathbf P^{\mathrm T}.
\label{eq:S-matrix-propagation}
\end{equation}
This formulation permits the input and output planes to have different spatial
sampling intervals and different numbers of samples while retaining a fully
differentiable propagation operator. No additional zero-padding, interpolation,
or explicit spatial-frequency band-limit filtering is applied; propagation is
evaluated directly between the specified input and output sampling grids.

The lens is modelled as a thin phase element. Although the implementation
supports both paraxial and spherical phase profiles, all results reported here
use the spherical form
\begin{equation}
\mathcal Q_f[E](x,y)
=
E(x,y)
\exp\!\left[
ik\left(
f-\sqrt{f^2+x^2+y^2}
\right)
\right].
\label{eq:S-lens}
\end{equation}
For $x^2+y^2\ll f^2$, this reduces to the usual paraxial thin-lens phase
\begin{equation}
\mathcal Q_f[E](x,y)
\simeq
E(x,y)
\exp\!\left[
-\frac{ik}{2f}(x^2+y^2)
\right].
\end{equation}
The lens is centred on the optical axis, with no transverse displacement in
the simulations presented here.

The spatial phase mask is modelled as a phase-only element,
\begin{equation}
\mathcal P_{\Phi}[E](x,y)
=
E(x,y)\exp[i\Phi(x,y)].
\label{eq:S-mask}
\end{equation}
The full simulated measurement operator is therefore
\begin{equation}
\Lop_{\Phi}
=
\mathcal F_{L_3}
\circ
\mathcal P_{\Phi}
\circ
\mathcal F_{L_2}
\circ
\mathcal Q_f
\circ
\mathcal F_{L_1},
\label{eq:S-full-operator}
\end{equation}
and the simulated camera measurement is
\begin{equation}
I(x,y)
=
\left|
\Lop_{\Phi}[E](x,y)
\right|^2.
\label{eq:S-camera-intensity}
\end{equation}

All optical fields and propagation matrices are represented using complex
single-precision tensors in the present implementation. The optical components
are applied sequentially.

\subsection{Spatial sampling and numerical parameters}

The numerical parameters used for the simulations are summarised in
Table~\ref{tab:S-parameters}. Distances and transverse coordinates are expressed
in micrometres.

\begin{table}[H]
\centering
\caption{Numerical parameters of the optical simulations that produced the optimised masks and all simulated reconstructions (ROI-constrained design, Supplementary Note~4), as set in the simulation code (\texttt{studies/run\_42\_roi\_constrained.py} and \texttt{studies/run\_direct\_reconstruction\_sweep.py} of the simulation repository). The experiment used 105\,mm and 60\,mm for $L_2$ and $L_3$ (Supplementary Note~4).}
\label{tab:S-parameters}
\begin{tabular}{lll}
\toprule
Parameter & Symbol & Value\\
\midrule
Wavelength & $\lambda$ & $0.6328\,\si{\micro\metre}$ (rounded to $0.633$ in the numerical model)\\
Fibre core radius & $r_{\mathrm{core}}$ & $12.5\,\si{\micro\metre}$ (42 LP modes)\\
Numerical aperture & $\mathrm{NA}$ & $0.10$\\
Cladding refractive index & $n_{\mathrm{clad}}$ & $1.46$\\
Fibre--lens distance & $L_1$ & $11.0\,\si{\milli\metre}$\\
Lens focal length & $f$ & $11.0\,\si{\milli\metre}$\\
Lens--mask distance & $L_2$ & $75.0\,\si{\milli\metre}$\\
Mask--camera distance & $L_3$ & $68.6\,\si{\milli\metre}$\\
Fibre grid and sampling & $N_0$, $\Delta x_0$ & $256\times256$, $\approx0.1973\,\si{\micro\metre}$\\
Lens grid and sampling & $N_1$, $\Delta x_1$ & $1024\times1024$, $4.0\,\si{\micro\metre}$\\
Phase plane & $N_2$, $\Delta x_2$ & $1024\times1024$, $8.0\,\si{\micro\metre}$ (SLM pitch)\\
Phase levels & & 256\\
Camera & $N_3$, $\Delta x_3$ & $1024\times1024$, $5.3\,\si{\micro\metre}$\\
\bottomrule
\end{tabular}
\end{table}

The fibre-plane sampling is determined by the numerical domain used to generate
the LP modes. In the present implementation the transverse domain extends from
$-R_{\mathrm{img}}$ to $+R_{\mathrm{img}}$, with
\begin{equation}
R_{\mathrm{img}}
=
r_{\mathrm{core}}+20\lambda.
\end{equation}
For the parameters in
Table~\ref{tab:S-parameters}, this gives
$R_{\mathrm{img}}=25.16\,\si{\micro\metre}$ and
$\Delta x_0=2R_{\mathrm{img}}/(256-1)\simeq0.1973\,\si{\micro\metre}$
for 256 samples including both endpoints.

\subsection{Power capture}

Each propagation maps between finite sampled planes, so power outside the camera window is not represented. Captured fractions are therefore referred to the power incident on the mask. For mode $j$ immediately before and after the final propagation, with square-grid pitches $\Delta x_{\mathrm{prev}}$ and $\Delta x_{\mathrm{cam}}$,
\begin{equation}
P_j^{\mathrm{prev}}=(\Delta x_{\mathrm{prev}})^2\sum_{m,n}|E^{\mathrm{prev}}_{jmn}|^2,\qquad
\widetilde E^{\mathrm{cam}}_{jmn}=\Delta x_{\mathrm{cam}}E^{\mathrm{cam}}_{jmn}.
\label{eq:S-discrete-power}
\end{equation}
The region of interest is a radially symmetric soft window with half weight at radius $R=2$ mm,
\begin{equation}
R_{\mathrm{px}}=\frac{1000R_{\mathrm{mm}}}{\Delta x_{\mathrm{cam}}},\qquad
W_{mn}=\exp\!\left[-\ln2\left(\frac{r_{mn}^2}{R_{\mathrm{px}}^2}\right)^p\right],\qquad p=4,
\label{eq:S-confinement-window}
\end{equation}
with $r_{mn}$ the pixel distance from the camera centre; for $\Delta x_{\mathrm{cam}}=5.3\,\si{\micro\metre}$, $R_{\mathrm{px}}\approx377.4$ pixels. The mean capture is
\begin{equation}
\bar\eta=\frac1M\sum_{j=1}^{M}\frac{\sum_{m,n}|\widetilde E^{\mathrm{cam}}_{jmn}|^2W_{mn}}{P_j^{\mathrm{prev}}},
\label{eq:S-confinement-loss}
\end{equation}
an average of per-mode ratios. The worst-case capture over all coherent superpositions, $\eta_{\min}$, is defined in Supplementary Note~4. The window enters only these constraints: it does not act as a physical aperture and does not mask the PMP matrix, whose condition number is evaluated on all camera pixels.

\section{Supplementary Note 4: Optimisation implementation}

\subsection{PMP construction}

The phase-mask optimisation is implemented entirely within the differentiable
PyTorch model described in Supplementary Note~3. The LP basis modes are first
propagated from the fibre plane to the SLM plane through the fibre--lens and
lens--SLM sections of the optical system. During optimisation, these fields are
then propagated through the trainable phase mask and from the SLM to the camera.

The calculated LP modes are normalised individually to unit discretised energy. For each mode $\Psi_j$,
\begin{equation}
    \sum_{\mathbf r}
    \Delta x_0^2
    \left|\Psi_j(\mathbf r)\right|^2
    =
    1,
    \qquad
    j=1,\ldots,M.
    \label{eq:S-mode-normalization}
\end{equation}
This normalisation is applied independently to every LP mode before propagation and before construction of the PMP matrix, ensuring that differences in the resulting detector signatures arise from the optical transformation and modal structure rather than from arbitrary differences in modal energy.

For a given phase mask $\Phi$, each basis mode produces a complex field at the
camera,
\begin{equation}
E_j(\mathbf r;\Phi)
=
\Lop_{\Phi}[\Psi_j](\mathbf r).
\label{eq:S-camera-mode-propagation}
\end{equation}
The camera fields are flattened spatially to form a matrix
$\mathbf A_{\Phi}$ whose $j$th column contains the detector-plane field
generated by mode $\Psi_j$. The pairwise-mode-product (PMP) matrix is then
constructed from all ordered pairs of modal fields,
\begin{equation}
\left[\mathbf{\Upsilon}_{\Phi}\right]_{\mathbf r,(i,j)}
=
E_i(\mathbf r;\Phi)
E_j^*(\mathbf r;\Phi),
\qquad
i,j=1,\ldots,M,
\label{eq:S-pmp-construction}
\end{equation}
so that $\mathbf{\Upsilon}_{\Phi}$ contains $M^2$ columns. The optimisation
therefore acts directly on the conditioning of the complete PMP measurement
space.

\subsection{Optimisation problem}

The mask minimises the PMP condition number subject to two power-capture constraints (main-text Eq.~5),
\begin{equation}
\min_{\phi}\ \kappa\!\left(\mathbf{\Upsilon}_{\Phi}\right)
\quad\text{subject to}\quad \bar\eta\geq0.98,\qquad\eta_{\min}\geq0.95,
\label{eq:S-optimization}
\end{equation}
with $\bar\eta$ from Eq.~\ref{eq:S-confinement-loss}. The second constraint bounds the captured fraction for every field in the 42-mode subspace, not only for the basis modes. For incident-mode fields collected in $B$ and camera fields collected in $A$ (including pixel-area weights), define $G_{\mathrm{in}}=B^{\mathrm H}B$ and $G_{\mathrm{ROI}}=A^{\mathrm H}WA$. Then $\eta_{\min}$ is the smallest eigenvalue of $G_{\mathrm{in}}^{-1/2}G_{\mathrm{ROI}}G_{\mathrm{in}}^{-1/2}$, the minimum captured fraction over coherent modal superpositions.

\subsection{Optimisation protocol}

\begin{table}[H]
\centering
\caption{Settings of the ROI-constrained optimisation.}
\label{tab:S-optimization-settings}
\begin{tabularx}{\linewidth}{@{}l>{\raggedright\arraybackslash}X@{}}
\toprule
Setting & Value\\
\midrule
Starts & 12 (seeds 12, 23, \ldots, 133 in steps of 11)\\
Initial phase & weak smooth random, RMS $\approx0.037$ rad after quantisation\\
Warm-up & 100 epochs, regularised trace-normalised $\log\det$ of the Gram matrix (ridge $10^{-3}$)\\
Main stage & 300 epochs minimising $\log\kappa$\\
Learning rate & 0.08\\
Conditioning pixels & 8,192 ROI pixels, resampled every epoch\\
Constraints & augmented Lagrangian (mean capture); projected steps with backtracking (both)\\
Checkpoint & lowest training condition number satisfying both constraints\\
\bottomrule
\end{tabularx}
\end{table}

The warm-up improves the whole singular-value spectrum before the condition number, which depends only on its extremes, is targeted. Resampling the conditioning pixels each epoch keeps the per-step cost low while the final condition numbers are computed on all 1,048,576 camera pixels (and, separately, on all 447,356 pixels inside the 2 mm circle). The optimisation schedule was chosen in pilot runs with seed 12; the reported ensemble contains seed 12 and eleven further starts, so it tests reproducibility across starts rather than providing an independent confirmation.

\subsection{Mask displayed in the SLM acquisition}

The SLM acquisition of Supplementary Note~6 displayed the ROI-constrained seed-12 mask (main-text Fig.~1b shows its L1-regularised version), exported without resampling as a $626\times626$ array of 256 phase levels at the $8\,\si{\micro\metre}$ SLM pitch (\texttt{figure1\_mask\_native\_uint8.npy}, SHA-256 \texttt{e082a47d\ldots}; decoding $\phi=-\pi+2\pi u/256$), i.e.\ the central $5.0\times5.0$\,mm of the designed phase plane. The SLM was placed 105\,mm from the collimating lens and the camera 60\,mm from the SLM, whereas the mask was designed for 75\,mm and 68.59\,mm (Supplementary Table~\ref{tab:S-parameters}). Because the camera-plane modal fields are calibrated from holography (Supplementary Note~6), reconstruction does not rely on the design model, and these differences change only the calibrated operator, not the procedure.

The absolute position of the beam on the SLM is not reproducible between alignments, so the mask was centred on the beam rather than at fixed SLM coordinates. Before the acquisition, the beam centre on the SLM was located with a single-spatial-mode beam taken directly from the laser through single-mode fibre: a circular aperture was displayed on the SLM and moved until the resulting diffraction pattern on the camera was circularly symmetric. The mask was then displayed centred at that position.

\subsection{Matched ROI-constrained phase-mask controls}

Fine pixel-scale random masks reach lower condition numbers (median 7.7) but retain only 42\% of the power in the ROI. Smooth random masks with comparable capture have a median condition number of 77, and the flat-phase SLM has 8,378. The optimised masks reduce the peak-to-mean intensity of the worst basis-mode launch from 234 for the flat-phase SLM to a median of 78. Peak loading was not part of the optimisation objective; this comparison concerns individual basis-mode launches, whereas the worst-state capture constraint applies to arbitrary coherent superpositions.
\label{sec:S-roi-controls}

The optimised masks are compared with flat and random controls in the same optical system, to test whether optimisation can jointly preserve power in a prescribed footprint, improve PMP conditioning and reduce peak detector loading. The comparison uses 42 LP modes at the rounded numerical wavelength of 633 nm (source wavelength 632.8 nm), core radius $12.5\,\mu$m and numerical aperture 0.10. The common phase-only plane is $1024\times1024$ at $8\,\mu$m pitch with 256 phase levels; the camera is $1024\times1024$ at $5.3\,\mu$m pitch, with area $29.4545\,\mathrm{mm}^2$. The final propagation distance is 68.5902 mm. These are the parameters of Supplementary Table~\ref{tab:S-parameters}; they describe the simulations, not the experimental acquisitions.

All masked systems use the same optics and incident-power normalisation. Controls comprise a flat-phase SLM, 144 tiled random masks (24 at each cell size 1, 2, 4, 8, 16 and 32 physical pixels), and 72 smooth random masks (12 at each phase scale 1, 1.5, 1.7, 2, 3 and 4). Smooth controls share the optimiser's $128\times128$ bilinear phase-control parameterisation and the same physical quantisation. A separate direct Fourier far-field reference is not equivalent to a flat-phase SLM in this system.

\section{Supplementary Note 5: Reconstruction across mode count and noise}

The full-sensor SNR range of $-15$ to 32 dB spans peak-to-noise ratios $I_{\max}/\sigma$ from about 3 to 700. At $-15$ dB, where the noise standard deviation is approximately one third of the brightest pixel, median coefficient NMSE is $4.8\times10^{-3}$ for 42 modes and $6.3\times10^{-3}$ for 51 modes, corresponding to relative $L^2$ errors of 6.9\% and 8.0\%. Independent noise is averaged over more than $10^{6}$ detector pixels while at most $51^2$ coefficient products are recovered. These results assume the specified independent-pixel noise and matched optical models.

The full-pixel direct-QR study in Fig.~4 of the main text uses ten physical core radii, fifteen realised-SNR targets from $-15$ to 32 dB and 1,000 independent fields per radius. All fields are reconstructed noiselessly and at each noise level with the unregularised least-squares estimator, one stored inverse per radius (for the five lowest SNR targets at 23 or more modes, the equivalent QR solve of the full PMP matrix on a GPU); fields are paired across noise levels, whereas detector-noise draws are independent. A single unchanged 42-mode-trained ROI mask is used throughout. The full $1024^2$ sensor is used for the inverse and the SNR, including pixels outside the 2 mm ROI. The ROI is an optical-design constraint, not a reconstruction crop. The source, mask hash, geometry, seeds, true/recovered coefficients and trial-level errors are supplied with the five low-SNR levels and the merged study at \path{reports/direct_reconstruction_statistics_lowsnr/} and \path{reports/direct_reconstruction_statistics_extended/}; copied summaries, all trial-level errors and validation are in \path{data/direct_statistics_20261022/}. Large inverse caches can be regenerated.

The reported NMSEs remove a unit-modulus global phase only. Field NMSE is evaluated with the sampled fibre-mode Gram matrix, preserving finite-grid nonorthogonality, and was cross-checked against explicit field images for displayed examples. Coefficient and field phase alignment are independent. Root NMSE is a relative $L^2$ error; it is not another NMSE. Every trial is retained. The optical forward model is shared by data generation and inversion, so these are matched-model noise results.

\subsection{Experimental inference speed}

Supplementary Table~\ref{tab:S-gpu-speed} gives the speed of the complete experimental pipeline (main-text Methods): a raw $1,024\times1,280$ frame is preprocessed to $256\times256$, multiplied by the stored $1,764\times65,536$ regularised pseudoinverse of the calibrated operator, and reduced to the 42 modal weights by the leading eigenpair of the $42\times42$ product matrix. The operators are the two reported 632.8 nm operators (Supplementary Note~6). For this matrix size the $42\times42$ eigendecomposition is faster on the CPU (batch-1 median 2.06--2.09 ms per frame in single precision) than on the GPU (2.59--2.62 ms), so the CPU is used. Synthesising the $64\times64$ field adds about 0.1 ms. In batches, throughput for frames held in GPU memory is largest at 64 frames; with host transfer it is limited by the upload of the raw frames. On all 500 held-out images of each acquisition the single-precision GPU result agrees with the reference implementation to NMSE $2.9\times10^{-11}$ (SLM) and $3.7\times10^{-9}$ (diffuser), and reproduces the eligible medians of 9.09\% and 4.61\%.

Storing the inverse in half precision shortens the stored-inverse product (Supplementary Table~\ref{tab:S-gpu-precision}). With fp16, the eligible median field NMSE changes by $-0.0003$ (SLM) and $-0.0004$ (diffuser) percentage points and the GPU result agrees with the reference to NMSE $\leq1.4\times10^{-7}$; bf16 deviates more ($\leq4.8\times10^{-6}$; median change up to $-0.003$ percentage points). No precision produced non-finite values. Single precision is the only one that reproduces both reported medians to two decimals and is used for all reported results.

The cost grows with the number of detector pixels. For a synthetic full-sensor input of $1,048,576$ pixels, the batch-1 median is 10.3 ms (97 frames/s) in single precision and 6.0 ms (167 frames/s) in fp16, still without iteration.

\begin{table}[!ht]
\centering
\caption{Single-image reconstruction speed on an NVIDIA GeForce RTX 3090 (raw frame to 42 modal weights, weights returned to host memory; single precision, eigendecomposition on the CPU). Batch 1: median and 95th percentile of 300 repetitions. Best batch: batch size with the largest throughput.}
\label{tab:S-gpu-speed}
\begin{tabular}{lcccc}
\toprule
 & \multicolumn{2}{c}{Batch 1} & GPU-resident frames & Frames from host\\
Operator & ms (95th pct.) & frames/s & best batch: frames/s & best batch: frames/s\\
\midrule
Diffuser ($\kappa=107$) & 2.09 (2.72) & 479 & 64: 4,015 & 32: 2,120\\
SLM ($\kappa=45$) & 2.06 (2.64) & 484 & 64: 4,162 & 16: 2,062\\
\bottomrule
\end{tabular}
\end{table}

\begin{table}[!ht]
\centering
\caption{Effect of the stored-inverse precision (batch 1, eigendecomposition on the CPU, without field synthesis; 500 held-out images per acquisition). Median change: eligible median field NMSE relative to single precision. Agreement: largest NMSE between GPU and reference reconstructions.}
\label{tab:S-gpu-precision}
\begin{tabular}{llcccc}
\toprule
Operator & Precision & ms (95th pct.) & frames/s & Median change (pp) & Agreement\\
\midrule
Diffuser & fp32 & 2.09 (2.72) & 479 & -- & $3.7\times10^{-9}$\\
 & fp16 & 1.85 (2.32) & 541 & $-0.0004$ & $1.4\times10^{-7}$\\
 & bf16 & 1.87 (2.56) & 534 & $-0.0035$ & $4.8\times10^{-6}$\\
SLM & fp32 & 2.06 (2.64) & 484 & -- & $2.9\times10^{-11}$\\
 & fp16 & 1.86 (2.54) & 538 & $-0.0003$ & $4.7\times10^{-8}$\\
 & bf16 & 1.88 (2.46) & 533 & $+0.0011$ & $2.4\times10^{-6}$\\
\bottomrule
\end{tabular}
\end{table}

\subsection{Full-sensor simulation operator}

A separate benchmark measures direct modal-coefficient recovery for the 42-mode simulation operator using every one of the $1024^2$ camera pixels. The complete $1764\times1048576$ complex64 inverse occupies 13.8 GiB and is resident before timing. Input is a calibrated intensity image in host RAM and output is the complex modal-coefficient vector in host RAM. Computation comprises inverse--intensity multiplication and rank-one positive-semidefinite projection, with the implementation's round-off symmetrisation. It includes no iterative coefficient optimisation or amplitude refit. It does not include synthesis of a near-field image from the recovered coefficients.

\begin{table}[!ht]
\centering
\caption{Resident-inverse single-image inference latency for 42 modes and all 1,048,576 camera pixels. Medians and 5th--95th percentiles summarise 30 timing repetitions after three warmups of the same image, not a distribution over independent optical states.}
\label{tab:S-inference-latency}
\begin{tabular}{lrr}
\toprule
Implementation & Median (ms) & 5th--95th percentiles (ms)\\
\midrule
CPU public decomposition API & 77.75 & 73.82--90.09\\
Metal multiplication + CPU projection & 53.67 & 53.17--56.63\\
\bottomrule
\end{tabular}
\end{table}

The CPU thread limit is 18. CPU timings use the public decomposition API, including input validation and real-to-complex conversion. Metal timing includes image upload, synchronisation and transfer of the small product estimate back to the CPU for eigendecomposition. All inverse entries remain simultaneously resident on Metal, partitioned into 14 row blocks for allocation compatibility. The real-packed layout stores adjacent real and imaginary rows and exploits real-valued intensity input without approximating or subsampling the inverse. Offline inverse construction, disk loading, initial inverse upload, optical propagation, camera acquisition, file I/O and display are excluded. These single-image latencies are not batched throughput or acquisition-to-display frame rates. Stage medians are independently measured and must not be summed to reproduce the separately timed API median.

The timed image is a saved simulated 13 dB state (realised SNR 12.9996 dB). The explicit-inverse product estimate agrees with the saved full-QR estimate to $7.87\times10^{-5}$ relative $L^2$, and the Metal/CPU rank-one product estimates differ by $1.80\times10^{-7}$ relatively. These are numerical agreement checks, not field-reconstruction errors. Recorded software versions are NumPy 2.5.2, SciPy 1.18.0 and PyTorch 2.13.0. Configuration, percentile timings and benchmark source are preserved in \path{data/inference_benchmark_20261016/}. The exact processor, installed memory and OS version were not recorded, limiting hardware-specific comparisons with this benchmark. This benchmark uses the full-camera simulated inverse, rather than the $256\times256$ experimental calibration operators.

The supplementary error maps show NMSE directly as a function of guided-mode count and intensity SNR. Median maps summarise typical performance, while the 5th and 95th percentiles show trial variability for each fixed physical basis and noise level.

\begin{figure}[H]\centering
\includegraphics[width=\linewidth]{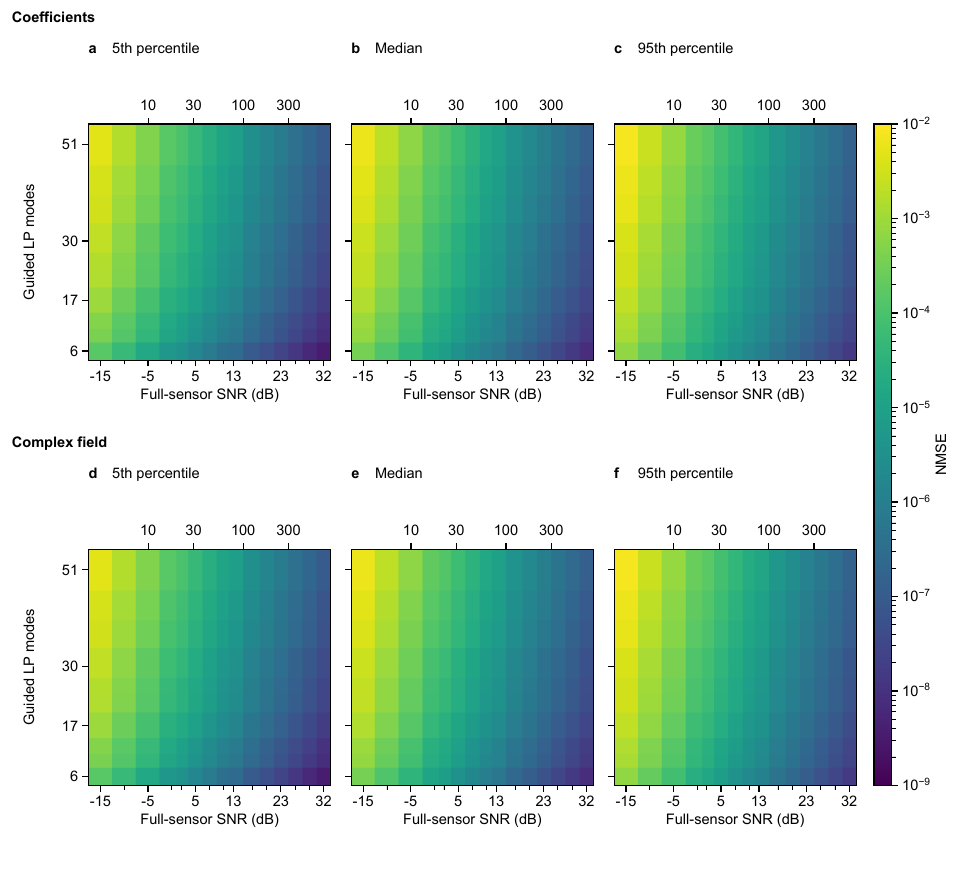}
\caption{NMSE versus guided LP mode count and noise level. The bottom axis gives the target full-sensor SNR and the top axis the equivalent peak-to-noise ratio $I_{\max}/\sigma$ (median offset 24.9 dB, main-text Methods); the vertical axis gives mode count; colour gives NMSE on one shared logarithmic scale. Panels (a--c) show modal-coefficient NMSE and (d--f) complex-field NMSE, each as the 5th percentile, median and 95th percentile, respectively. Each cell statistic uses all 1,000 trials for that mode-count/SNR condition; percentile ranges describe trial variability, not confidence intervals. Cells are centred at the fifteen tested SNRs ($-15$, $-10$, $-5$, 0, 2, 5, 8, 11, 13, 17, 20, 23, 26, 29 and 32 dB) and the ten tested mode counts. Cell edges lie midway between adjacent tested coordinates; no interpolation or smoothing is performed. Changing mode count changes the physical core radius and basis. Fields are paired across SNR levels; coefficient and field errors use their independently optimal global-phase alignments.}
\label{fig:S-noise-curve}
\end{figure}

The 10-bit camera model uses 128 ADU pedestal and 700 ADU peak signal per field, with Gaussian noise calibrated to each requested realised SNR. At the five targets below 5 dB the noise standard deviation (medians of about 30--220 ADU) approaches or exceeds the pedestal; these levels are therefore simulated without range clipping, so that the noise stays Gaussian (a 10-bit ADC would have clipped a median 27\% of the pixels at $-15$ dB and 0.05\% at 0 dB). At the other targets clipping is applied but affects at most $1.1\times10^{-5}$ of the pixels. Over the whole range the coefficient NMSE is approximately inversely proportional to the linear SNR: from $-15$ to 32 dB its median falls by a factor of $(4.6$--$5.1)\times10^{4}$ for the individual mode counts, matching the factor $5.0\times10^{4}$ (47 dB) in SNR. There is no electron/ADU calibration and no shot noise. Per-field gain/noise adjustment supports equal-SNR comparison but does not predict errors at fixed exposure. Independent noise across over a million pixels is strongly averaged by the low-dimensional estimator; correlated detector noise and optical mismatch require separate tests. The 5th--95th percentile ranges describe trial variability for this fixed mask, not confidence intervals over training seeds or optical hardware.

\begin{figure}[H]\centering
\includegraphics[width=\linewidth]{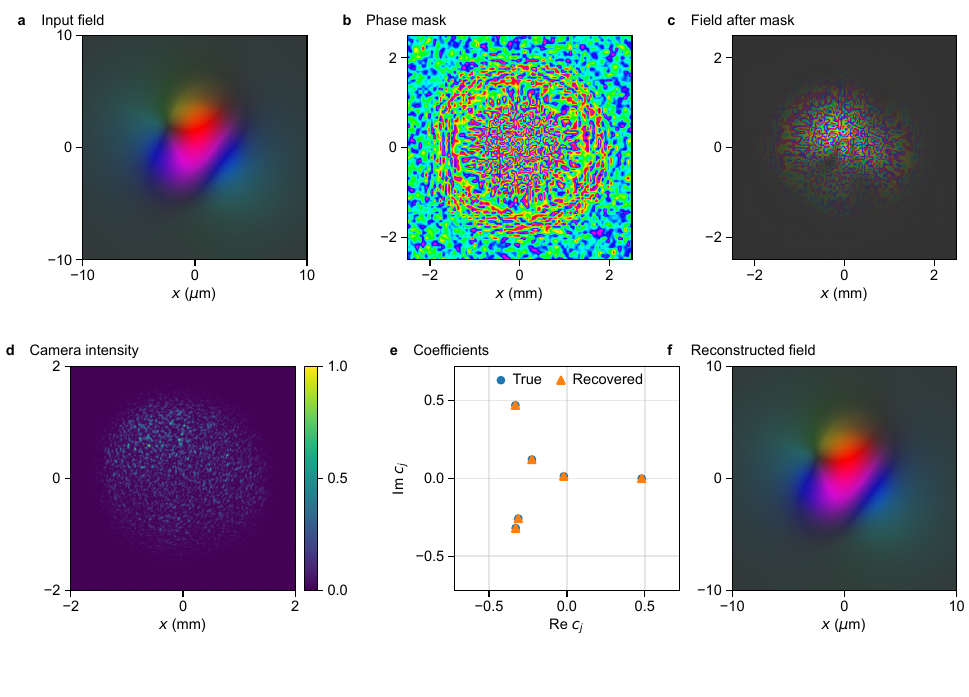}
\includegraphics[width=0.28\linewidth]{figures/display_keys/field_phase.pdf}
\caption{6 guided scalar LP modes; measured full-sensor SNR 12.993 dB. Trial 205 of 1,000 was selected by coefficient NMSE closest to the ensemble median. Coefficient NMSE $5.867\times10^{-7}$ (percentile 50.1); complex-field NMSE $5.866\times10^{-7}$ (percentile 50.1). The six panels show input field, phase mask, field after the mask, noisy camera intensity, true and retrieved coefficients, and reconstructed field. Hue encodes phase and brightness amplitude; true and reconstructed fields share an amplitude scale. Camera normalisation and displayed crops are for visualisation only; all sensor pixels enter inference.}
\label{fig:S-median-6m-13}
\end{figure}

\begin{figure}[H]\centering
\includegraphics[width=\linewidth]{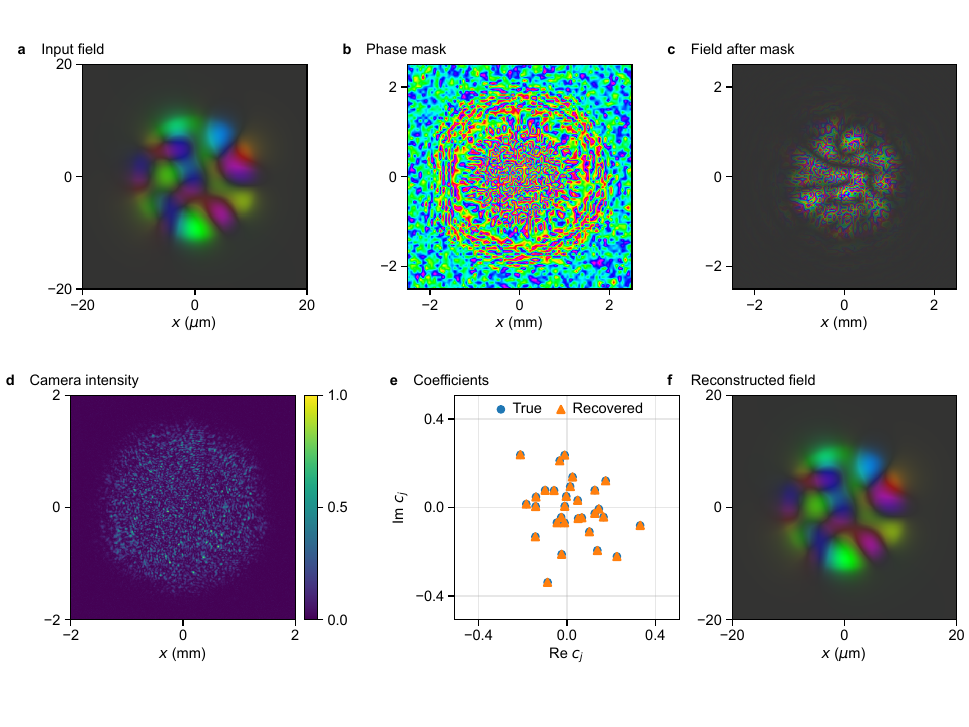}
\caption{Direct full-PMP reconstruction with the fixed 42-mode-trained phase mask. 30 guided scalar LP modes; measured full-sensor SNR 12.993 dB. Trial 596 of 1,000 was selected by coefficient NMSE closest to the ensemble median. Coefficient NMSE $4.381\times10^{-6}$ (percentile 50.1); complex-field NMSE $4.382\times10^{-6}$ (percentile 50.1).  The six panels show input field, phase mask, field after the mask, noisy camera intensity, true and retrieved coefficients, and reconstructed field. Hue encodes phase and brightness amplitude; true and reconstructed fields share an amplitude scale. Camera normalisation and displayed crops are for visualisation only; all sensor pixels enter inference.}
\label{fig:S-median-30m}
\end{figure}

Supplementary Figs.~\ref{fig:S-median-6m-13}--\ref{fig:S-median-51m-17} show, for four mode counts at 13 and 17 dB, the trial whose coefficient NMSE is closest to the ensemble median (\texttt{reporting/build\_median\_examples.py}). The camera images, input and reconstructed fields, and coefficients are read directly from the original saved study arrays. The reported trial, SNR and errors are those of the study; no new noise is drawn when regenerating these panels.

\begin{figure}[H]\centering
\includegraphics[width=\linewidth]{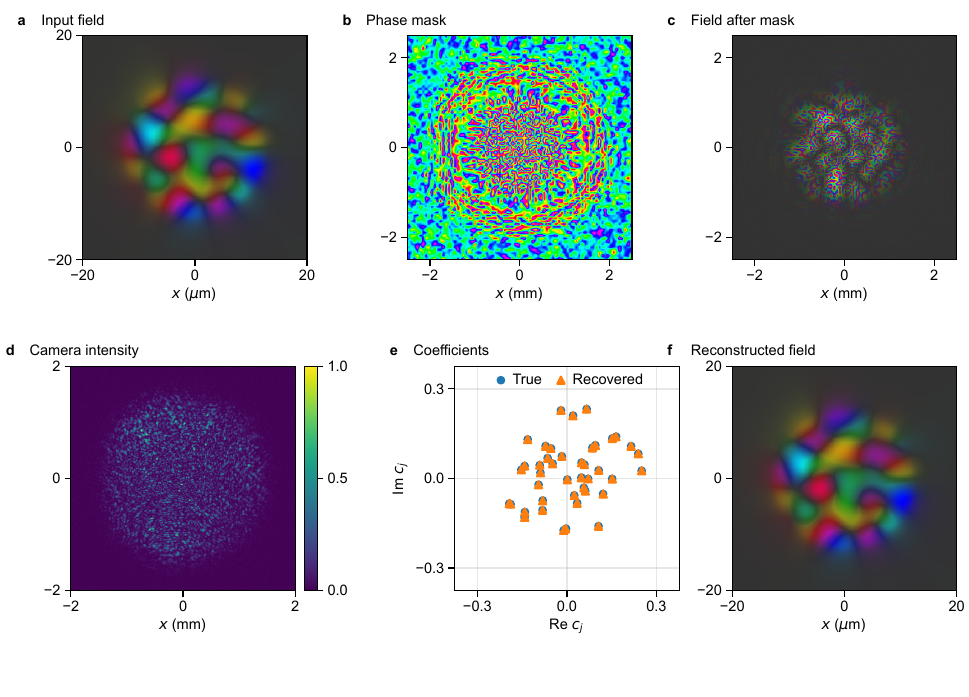}
\caption{42 guided scalar LP modes; measured full-sensor SNR 12.997 dB. Trial 306 of 1,000 was selected by coefficient NMSE closest to the ensemble median. Coefficient NMSE $7.613\times10^{-6}$ (percentile 50.0); complex-field NMSE $7.614\times10^{-6}$ (percentile 50.0). The six panels show input field, phase mask, field after the mask, noisy camera intensity, true and retrieved coefficients, and reconstructed field. Hue encodes phase and brightness amplitude; true and reconstructed fields share an amplitude scale. Camera normalisation and displayed crops are for visualisation only; all sensor pixels enter inference.}
\end{figure}

\begin{figure}[H]\centering
\includegraphics[width=\linewidth]{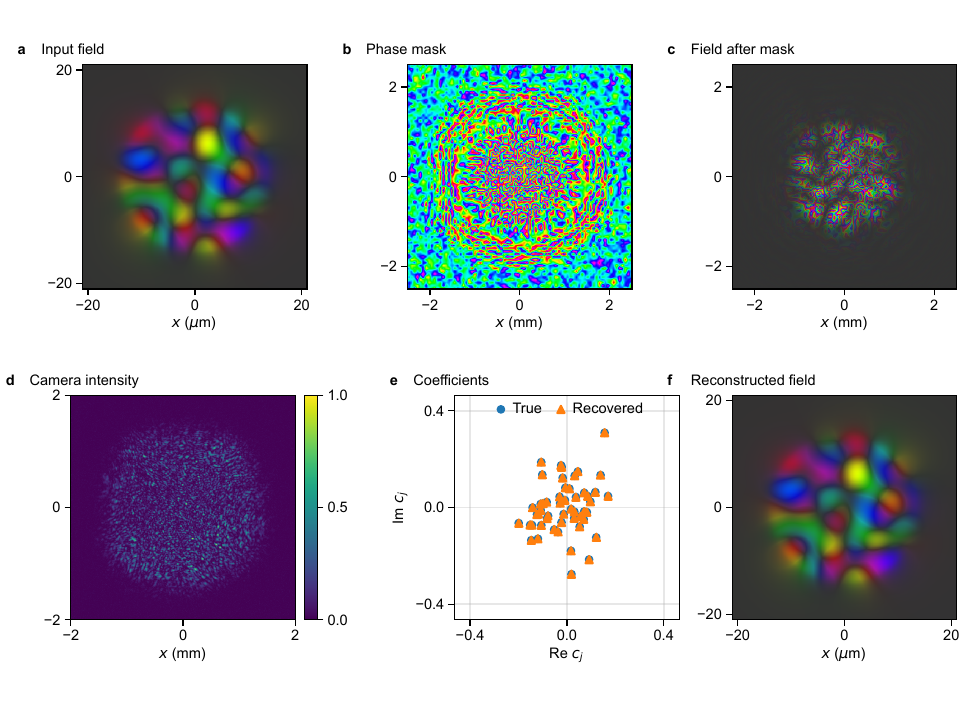}
\caption{51 guided scalar LP modes; measured full-sensor SNR 13.006 dB. Trial 760 of 1,000 was selected by coefficient NMSE closest to the ensemble median. Coefficient NMSE $1.007\times10^{-5}$ (percentile 50.0); complex-field NMSE $1.007\times10^{-5}$ (percentile 50.0). The six panels show input field, phase mask, field after the mask, noisy camera intensity, true and retrieved coefficients, and reconstructed field. Hue encodes phase and brightness amplitude; true and reconstructed fields share an amplitude scale. Camera normalisation and displayed crops are for visualisation only; all sensor pixels enter inference.}
\end{figure}

\begin{figure}[H]\centering
\includegraphics[width=\linewidth]{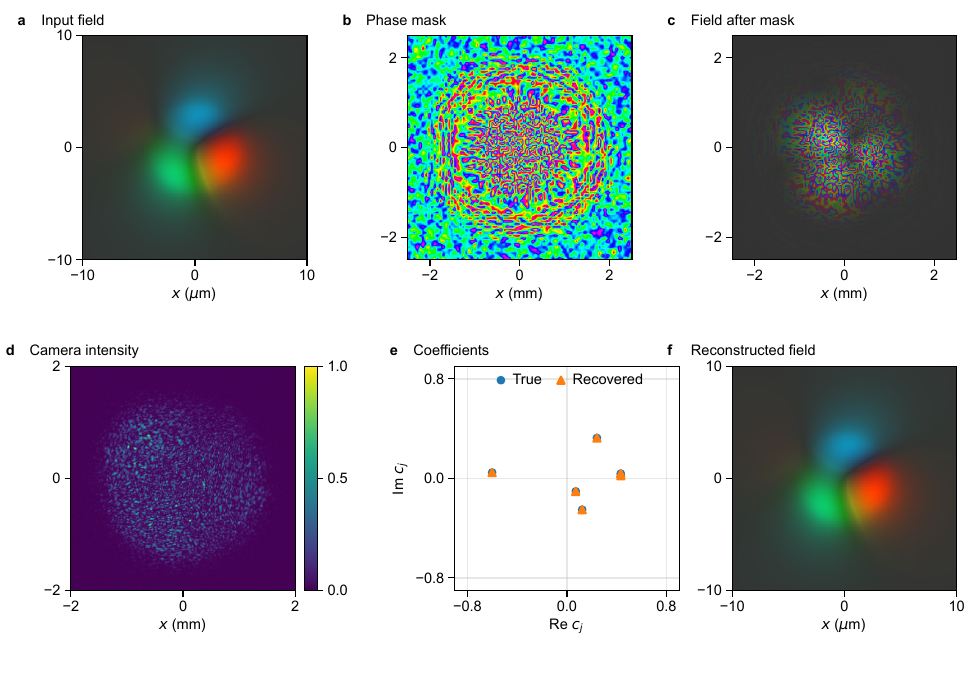}
\caption{6 guided scalar LP modes; measured full-sensor SNR 16.994 dB. Trial 39 of 1,000 was selected by coefficient NMSE closest to the ensemble median. Coefficient NMSE $2.239\times10^{-7}$ (percentile 50.0); complex-field NMSE $2.239\times10^{-7}$ (percentile 50.0). The six panels show input field, phase mask, field after the mask, noisy camera intensity, true and retrieved coefficients, and reconstructed field. Hue encodes phase and brightness amplitude; true and reconstructed fields share an amplitude scale. Camera normalisation and displayed crops are for visualisation only; all sensor pixels enter inference.}
\end{figure}

\begin{figure}[H]\centering
\includegraphics[width=\linewidth]{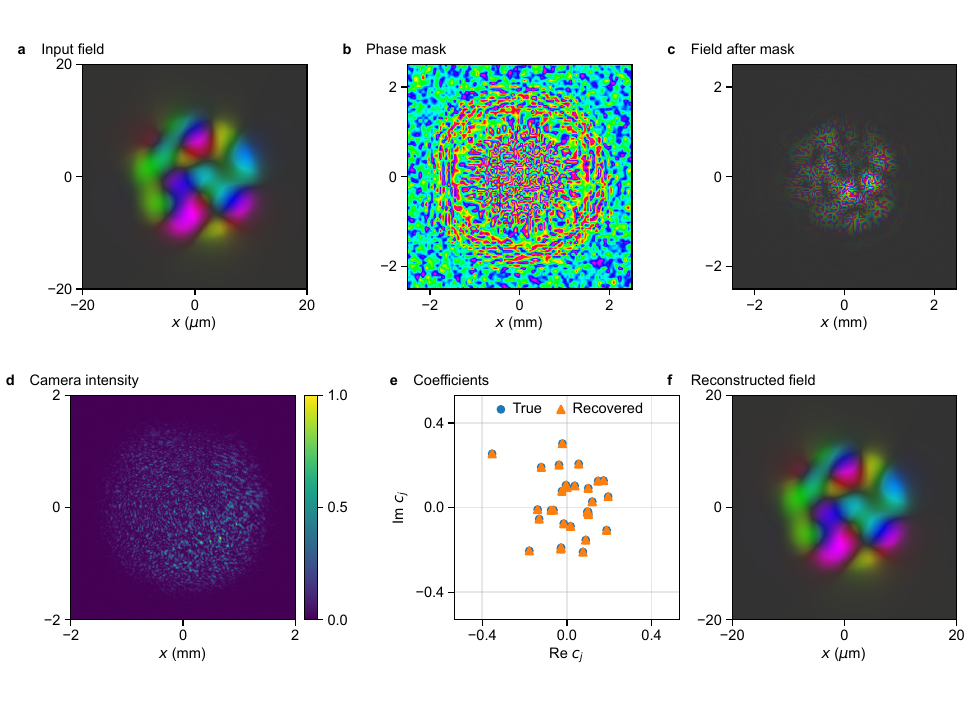}
\caption{30 guided scalar LP modes; measured full-sensor SNR 16.992 dB. Trial 969 of 1,000 was selected by coefficient NMSE closest to the ensemble median. Coefficient NMSE $1.760\times10^{-6}$ (percentile 50.0); complex-field NMSE $1.760\times10^{-6}$ (percentile 49.9). The six panels show input field, phase mask, field after the mask, noisy camera intensity, true and retrieved coefficients, and reconstructed field. Hue encodes phase and brightness amplitude; true and reconstructed fields share an amplitude scale. Camera normalisation and displayed crops are for visualisation only; all sensor pixels enter inference.}
\end{figure}

\begin{figure}[H]\centering
\includegraphics[width=\linewidth]{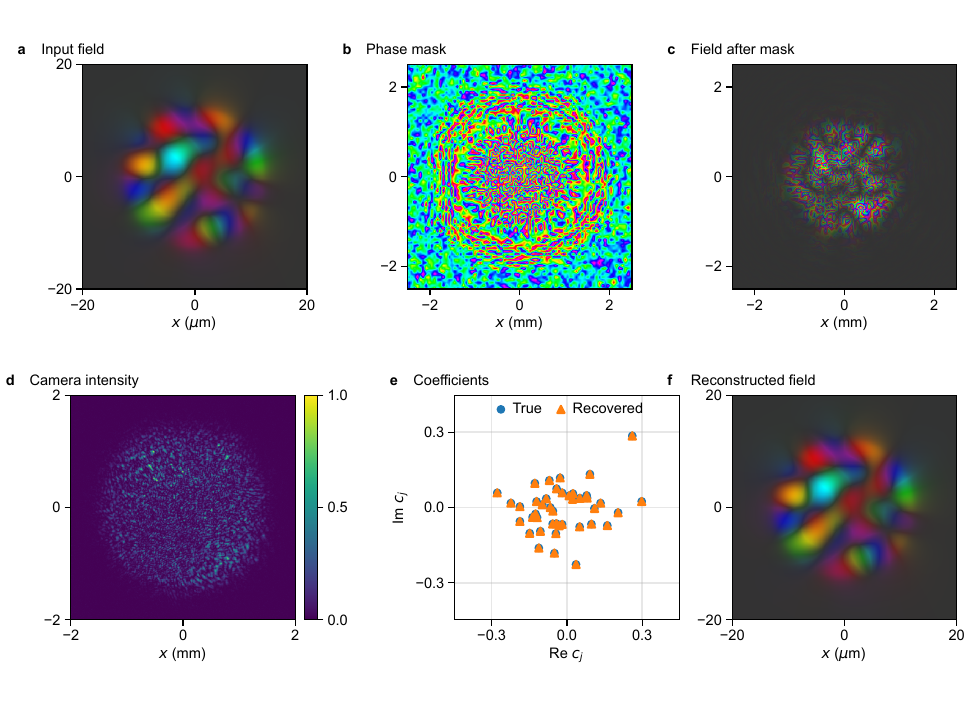}
\caption{42 guided scalar LP modes; measured full-sensor SNR 16.994 dB. Trial 125 of 1,000 was selected by coefficient NMSE closest to the ensemble median. Coefficient NMSE $3.038\times10^{-6}$ (percentile 50.0); complex-field NMSE $3.038\times10^{-6}$ (percentile 50.0). The six panels show input field, phase mask, field after the mask, noisy camera intensity, true and retrieved coefficients, and reconstructed field. Hue encodes phase and brightness amplitude; true and reconstructed fields share an amplitude scale. Camera normalisation and displayed crops are for visualisation only; all sensor pixels enter inference.}
\end{figure}

\begin{figure}[H]\centering
\includegraphics[width=\linewidth]{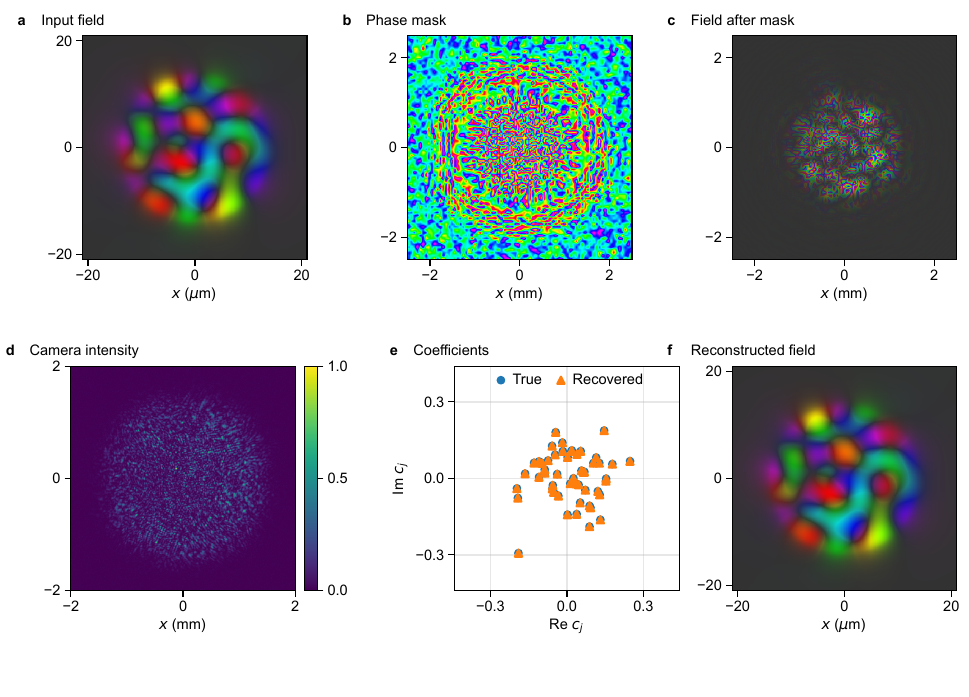}
\caption{51 guided scalar LP modes; measured full-sensor SNR 16.999 dB. Trial 5 of 1,000 was selected by coefficient NMSE closest to the ensemble median. Coefficient NMSE $3.993\times10^{-6}$ (percentile 50.1); complex-field NMSE $3.993\times10^{-6}$ (percentile 50.1). The six panels show input field, phase mask, field after the mask, noisy camera intensity, true and retrieved coefficients, and reconstructed field. Hue encodes phase and brightness amplitude; true and reconstructed fields share an amplitude scale. Camera normalisation and displayed crops are for visualisation only; all sensor pixels enter inference.}
\label{fig:S-median-51m-17}
\end{figure}

\section{Supplementary Note 6: Experimental reconstruction}
\label{sec:S-experiments}

\subsection{Data acquisition and calibration split}

The two cases are diffuser glass, acquisition \path{rare_mode_targeted_20261001_173945} (created 1 October 2026), and SLM displaying the optimised seed-12 mask (Supplementary Note~4), acquisition \path{rare_mode_targeted_figure1_native_20261018_234905} (created 18 October 2026).

Both cases use the same 42-component LP synthesis basis of the FG025LJA fibre, computed at the source wavelength of 632.8 nm, $256\times256$ processed symmetry-broken-plane intensities and $64\times64$ complex near fields. Each contains 5,000 measurements from 250 launch groups and 20 controller states per group: symmetry-broken-plane intensity, hologram and near-field intensity for the diffuser acquisition, and symmetry-broken-plane intensity and hologram for the SLM acquisition. Every tenth launch group, 25 in all (500 measurements), is held out, leaving 4,500 for calibration; no launch group crosses the split. Experimental conditions and calibration objectives differ across cases.

The source was a 632.8\,nm FBG-stabilised laser diode. Its output was linearly polarised and coupled through approximately 2\,m of polarisation-maintaining single-mode fibre. The few-mode fibre (FMF; SMF-28e) was approximately 2\,m long and was driven through a motorised polarisation controller (Thorlabs MPC320), which varied the launched modal superposition and scrambled the polarisation. The FMF was butt-coupled across a free-space gap to 4\,m of multimode fibre (MMF; Thorlabs FG025LJA); the fibres were not spliced. Their ends were held in the V-groove clamps of a Fujikura 70S fusion splicer in manual-alignment mode, with no arc applied. Its microscope and three-axis stages set the transverse core offset and axial gap described in Methods.

The MMF output was collimated by lens L1 (Thorlabs C220TME, nominal focal length 11\,mm). The polarisation was scrambled in the MMF and then selected to a single linear polarisation in free space by a film polariser (Thorlabs LPVIS050-MP2) after L1; the field therefore remained linearly polarised through the subsequent free-space optics. A cube beamsplitter divided the light between the reconstruction and holography channels. In the SLM configuration, the reflective phase-only SLM (HOLOEYE ERIS-1.1) was 105\,mm from L1, and camera C1 (Ximea MQ013RG-E2) was 60\,mm from the SLM. Camera C2 (Thorlabs LP126MU) recorded the near-field holograms through lens L2 (Thorlabs AC254-500-A, nominal focal length 500\,mm); the reference arm was collimated with a Thorlabs F810FC-635 fibre collimator and combined with the signal at a non-collinear angle. In the diffuser configuration, the SLM was removed and a ground-glass diffuser (Thorlabs DG10-220-MD, 220-grit) was placed approximately 2\,cm in front of C1. The remaining optical arrangement was unchanged. The mask-design distances (75\,mm from L1 to mask and 68.59\,mm from mask to camera) differ from the experimental SLM distances above; the experimental operator was calibrated from holography. The apparatus and spacings are schematic, and the diffuser--camera distance is approximate.

\subsection{Preprocessing and reference eligibility}

Per-frame exposure metadata select or linearly interpolate embedded dark mean frames. Corrected intensity is clipped at zero, cropped and normalised by its spatial mean before pooling. The mixing-camera crop uses rows 0--1023 and columns 128--1151 (zero-based), followed by $4\times4$ mean pooling. Diffuser near-camera frames of $520\times512$ pixels use rows 4--515; SLM frames already have $512\times512$ pixels.

Fourier-sideband extraction uses carrier-bin rolls $(52,163)$ for the diffuser and $(41,211)$ for the SLM acquisition. The mask radius is 48 bins with a five-bin cosine taper. The inverse transform is root-mean-square normalised, pooled in $8\times8$ blocks and divided by a fixed linear phase factor fitted on calibration references. No shared quadratic correction is applied.

Any raw hologram science-region pixel at or above 4,094 DN, or any non-finite pixel, excludes that reference before shared phase fitting and training. Additional reference-quality and LP-representability checks, with thresholds fitted on calibration references only, are recorded by the fixed eligibility mask.

\begin{table}[!ht]
\centering
\caption{Measurements retained after hologram quality screening. Eligible references additionally pass quality and LP checks. Counts are measurements (trios or duos).}
\label{tab:S-experimental-eligibility}
\begin{tabular}{lrr}
\toprule
Population & Diffuser & SLM\\
\midrule
Acquired & 5,000 & 5,000\\
Clipped acquired holograms & 1,250 & 1,741\\
Calibration & 4,500 & 4,500\\
Clipped calibration holograms & 1,124 & 1,588\\
Unsaturated calibration & 3,376 & 2,912\\
Eligible calibration & 3,296 & 1,007\\
Held-out & 500 & 500\\
Clipped held-out holograms & 126 & 153\\
Unsaturated held-out & 374 & 347\\
Eligible held-out & 368 & 150\\
\bottomrule
\end{tabular}
\end{table}

In the SLM acquisition, many unsaturated holograms are dominated by noise; its eligible held-out fields come from 19 of the 25 held-out launch groups. Its auto-exposure diagnostics do not consistently describe the saved frames (1,130 saved holograms are saturated despite a zero quality flag), so saturation was tested on the raw saved pixels only. Sparse clipping of the symmetry-broken-plane camera is retained (in the SLM acquisition, 3,009 frames with at most $9\times10^{-5}$ of the crop clipped).

\subsection{Reference quality and excluded measurements}

The exclusions originate in the holographic reference channel. To keep both cameras below saturation, the laser diode was operated just above its lasing threshold. Some holograms had weak or absent interference fringes, consistent with instability of the source coherence near threshold. The symmetry-broken-plane image used for reconstruction does not depend on interference with the reference arm. Of the 500 held-out measurements of the SLM acquisition, 350 are not eligible: 153 have saturated hologram pixels, and 197 are unsaturated but fail the reference-quality checks.

\begin{figure}[!ht]
\centering
\includegraphics[width=0.75\linewidth]{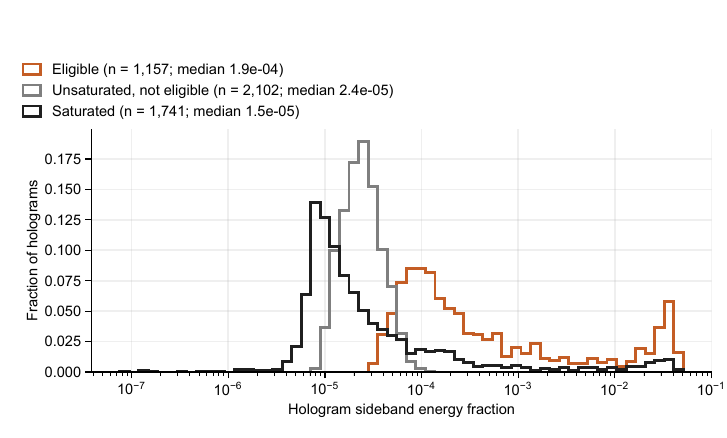}
\caption{Fringe energy of all 5,000 holograms of the SLM acquisition (calibration and held-out): fraction of the hologram's Fourier energy in the tapered off-axis sideband, for eligible references, unsaturated references that fail the quality checks, and holograms with saturated pixels. Each group is normalised to its number of holograms; logarithmic bins. Sideband energy does not depend on the LP model; eligibility is that of the 632.8 nm processing.}
\label{fig:S-fringe-energy}
\end{figure}

Supplementary Fig.~\ref{fig:S-fringe-energy} gives the fraction of hologram energy in the off-axis sideband for all 5,000 SLM holograms. Eligible holograms have a median sideband fraction of $1.9\times10^{-4}$, whereas unsaturated non-eligible holograms have $2.4\times10^{-5}$, eight times less; 91\% of the latter lie below the 5th percentile of the eligible holograms. Most saturated holograms ($1.5\times10^{-5}$) also have weak fringes. Low fringe energy, rather than saturation alone, therefore characterises the excluded references.

\begin{figure}[!ht]
\centering
\includegraphics[width=\linewidth]{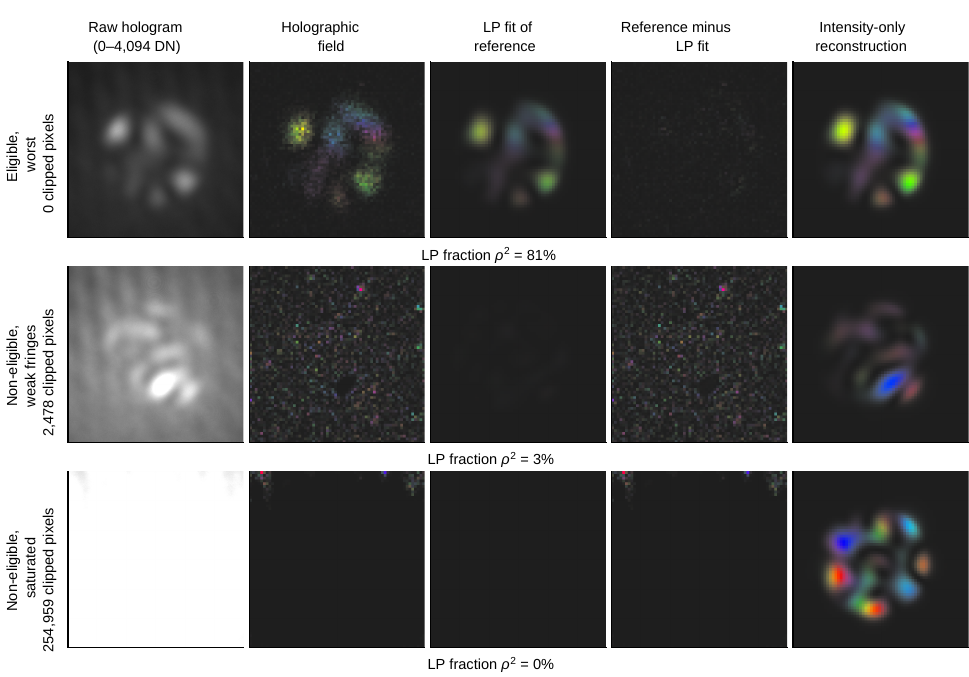}
\caption{Holographic references of the SLM acquisition. Rows: the worst eligible held-out field (\texttt{launch\_149\_mpc\_005}; field NMSE 19.8\%); a non-eligible held-out field with weak fringes (\texttt{launch\_159\_mpc\_016}; 1\% of hologram pixels clipped); a non-eligible held-out field with a saturated hologram (\texttt{launch\_189\_mpc\_015}; 97\% clipped). Columns: raw $512\times512$ hologram (0--4,094 DN); extracted and phase-corrected reference field; its least-squares fit in the fixed 42-mode LP basis, with the LP fraction $\rho^2$; reference minus fit; field reconstructed from the symmetry-broken-plane intensity alone. Reference, fit and remainder share the reference peak-intensity scale; the reconstruction is shown on its own intensity scale, with only its global phase aligned to the fit. The non-eligible raw holograms are from a separate diagnostic export and are not part of any reported score.}
\label{fig:S-reference-examples}
\end{figure}

Supplementary Fig.~\ref{fig:S-reference-examples} shows the consequence. The worst eligible reference still carries clear fringes, but its extracted field is noisy: only 81\% of its energy lies in the 42-mode LP subspace, and the remaining 19\% is noise and light outside the modelled subspace. Its field NMSE is 19.8\%; relative to the LP fit of the reference, the intensity-only reconstruction has an error of 1.0\%. In a non-eligible hologram with weak fringes, the extracted field has only a small LP component (3\% LP fraction), and in a saturated one no field can be extracted. The intensity-only reconstructions of these fields are well-formed modal fields, but without a valid reference their accuracy cannot be assessed, so they are excluded from all reported scores. For completeness, Supplementary Table~\ref{tab:S-experimental-metrics} lists all-held-out statistics, which include these invalid references; relative to the LP projection of each reference, the median error over all 347 unsaturated held-out SLM fields is still 1.3\%.

\subsection{Calibration and reconstruction}

Camera-plane modal fields were calibrated by optimisation using holography-derived modal labels. The calibrated operator contains $U\in\mathbb C^{42\times65536}$, LP synthesis fields $L\in\mathbb C^{42\times64\times64}$, a fixed phase factor, regularised inverse Cholesky factor and unregularised Gram spectrum. These quantities remain fixed during held-out inference.

With zero-based component-pair indexing,
\begin{equation}
P_{p,iM+j}=U_{i,p}\overline{U_{j,p}},\quad M=42,\qquad G=P^{\mathrm H}P.
\end{equation}
This factorisation specifies the full $65536\times1764$ PMP. The inverse computes
\begin{equation}
\widehat q=(G+\lambda_{\mathrm r}I)^{-1}P^{\mathrm H}i_d,\qquad
\lambda_{\mathrm r}=10^{-3}\operatorname{tr}(G)/1764.
\end{equation}
After reshaping to $Q$, the recovered product matrix is Hermitian by construction. This is followed by eigenvalue extraction and field reconstruction:
\begin{equation}
\widehat a=\sqrt{\lambda_1}\,v_1,\qquad
\widehat E=\sum_{i=1}^{42}\widehat a_iL_i.
\end{equation}

\subsection{Conditioning and metrics}

For the eligible SLM references, the component outside the 42-mode subspace accounts for a median NMSE of 8.4\% and predicts each sample's field error with correlation 0.998. Against the 42-mode projection of the reference, the reconstruction has median NMSE 0.7\% for the SLM (5th--95th percentiles 0.4--1.3\%) and 2.0\% for the diffuser, equal to the modal-coefficient errors. The SLM therefore has the smaller modal-subspace error despite its larger reference-field error. Averaged over the eligible SLM test fields, all 42 reconstructed LP weights agree with the holographic labels within 0.24 percentage points, including 14 components carrying less than 1\% of the normalised weight (main-text Fig.~6c).

Reported conditioning is
\begin{equation}
\kappa(P)=\sqrt{\lambda_{\max}(G)/\lambda_{\min}^{\mathrm{retained}}(G)},\qquad
\lambda_j(G)>2^{-23}\lambda_{\max}(G).
\end{equation}
All 1,764 eigenvalues pass this float32 cutoff. Reported values are 106.79 (diffuser) and 45.47 (SLM); values recomputed from the exported operators are 106.82 and 45.47.

Field and coefficient errors use optimal global complex-scalar alignment as defined in the main Methods. Coefficient references are holographic LP projections, whereas field references are the full corrected holographic fields. Per sample, NMSE equals $1-\gamma^2$; this relation cannot be applied directly to the mean coherence. Medians use the lower-middle observation, and percentiles use linear interpolation.

\begin{table}[!ht]
\centering
\caption{Experimental reconstruction metrics at $256\times256$ detector sampling and $64\times64$ field sampling, with the LP basis at 632.8 nm. Eligible statistics use test fields with valid holographic references; all-held-out statistics also include invalid references. Field NMSE and coherence remove an optimal global complex scalar (main-text Methods). Coefficient errors compare the reconstruction with the least-squares LP projection of the reference; they equal the NMSE of the 42 complex LP coefficients. Each test set contains 25 launch groups with correlated controller states; percentiles describe the empirical population, not confidence intervals.}
\label{tab:S-experimental-metrics}
\begin{tabular}{lcc}
\toprule
Metric & Diffuser & SLM\\
\midrule
Eligible test fields / all test fields & 368 / 500 & 150 / 500\\
Eligible field NMSE, mean (\%) & 5.00 & 9.71\\
Eligible field NMSE, median (\%) & 4.61 & 9.09\\
Eligible field NMSE, 5th--95th (\%) & 3.28--8.19 & 4.15--18.15\\
Eligible field NMSE, minimum--maximum (\%) & 2.54--27.17 & 2.40--19.78\\
Eligible field coherence, mean & 0.9746 & 0.9499\\
Eligible field coherence, median & 0.9765 & 0.9530\\
Reference out-of-subspace fraction $1-\rho^2$, median (\%) & 2.56 & 8.44\\
Eligible coefficient NMSE, mean (\%) & 2.38 & 0.73\\
Eligible coefficient NMSE, median (\%) & 2.01 & 0.66\\
Eligible coefficient NMSE, 5th--95th (\%) & 0.94--4.78 & 0.39--1.27\\
Eligible coefficient coherence, mean & 0.9880 & 0.9964\\
All-held-out field NMSE, mean (\%) & 6.50 & 41.52\\
All-held-out field NMSE, median (\%) & 4.94 & 35.94\\
All-held-out field NMSE, 5th--95th (\%) & 3.29--13.25 & 4.86--92.92\\
All-held-out field coherence, mean & 0.9662 & 0.7264\\
Unregularised PMP condition number & 106.79 & 45.47\\
\bottomrule
\end{tabular}
\end{table}

The field NMSE of each eligible SLM reconstruction is almost entirely explained by the out-of-subspace fraction of its reference (correlation 0.998 across the 150 fields): relative to the LP projection of the reference, the error never exceeds 1.8\%. The dependence is weaker for the diffuser (correlation 0.61), whose references are closer to the LP subspace. Over all 347 unsaturated held-out references of the SLM acquisition, irrespective of the remaining quality checks, the mean field NMSE is 36.1\% but the median error relative to the LP projection is 1.3\%, but valid references are required to assess reconstruction accuracy.

\begin{figure}[!ht]
\centering
\includegraphics[width=0.55\linewidth]{figures/experimental_632p8nm/summary_distribution.pdf}\hfill
\includegraphics[width=0.40\linewidth]{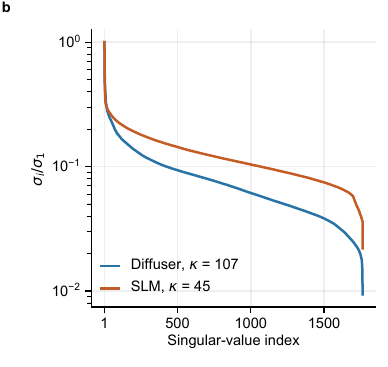}
\caption{Left: distributions of the complex-field NMSE over the eligible test fields (filled) and of the error relative to the 42-mode projection of each holographic reference (dotted), for the diffuser (368 fields) and SLM (150 fields) acquisitions; bin width 0.5\%, each case normalised to its number of fields. Right: normalised singular values of the two calibrated PMP matrices.}
\label{fig:S-nmse-distribution}
\end{figure}

Modal-content plots first normalise each sample's coefficient powers to unit sum, then average eligible holographic measurements. They document modal coverage, not component-wise reconstruction accuracy.

\begin{figure}[p]
\centering
\includegraphics[width=\linewidth]{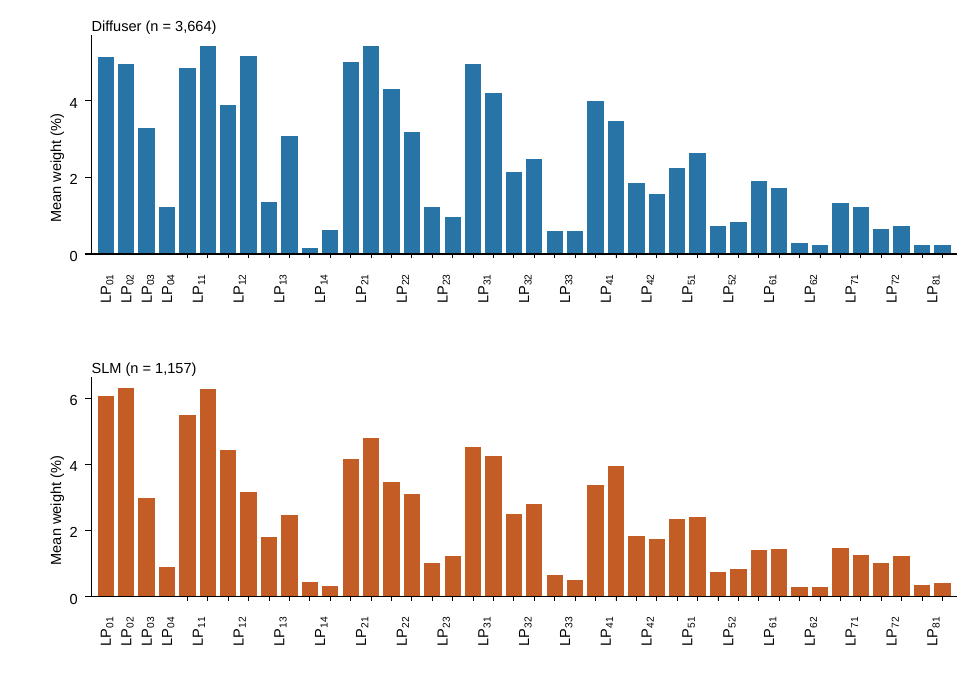}
\caption{Mean normalised holographic LP coefficient weights in the eligible populations (calibration and held-out) of the diffuser and SLM acquisitions. Each sample is normalised before averaging. These are coefficient weights in the fitted LP basis, not calibrated optical powers.}
\label{fig:S-experimental-modal}
\end{figure}
\clearpage

\subsection{Selected examples}

Five lowest- and five highest-NMSE eligible test fields are selected per case using the held-out scores. The examples are selected extremes. Hue represents phase and saturation/brightness represents intensity.

\begin{figure}[!h]
\centering
\includegraphics[width=0.9\linewidth]{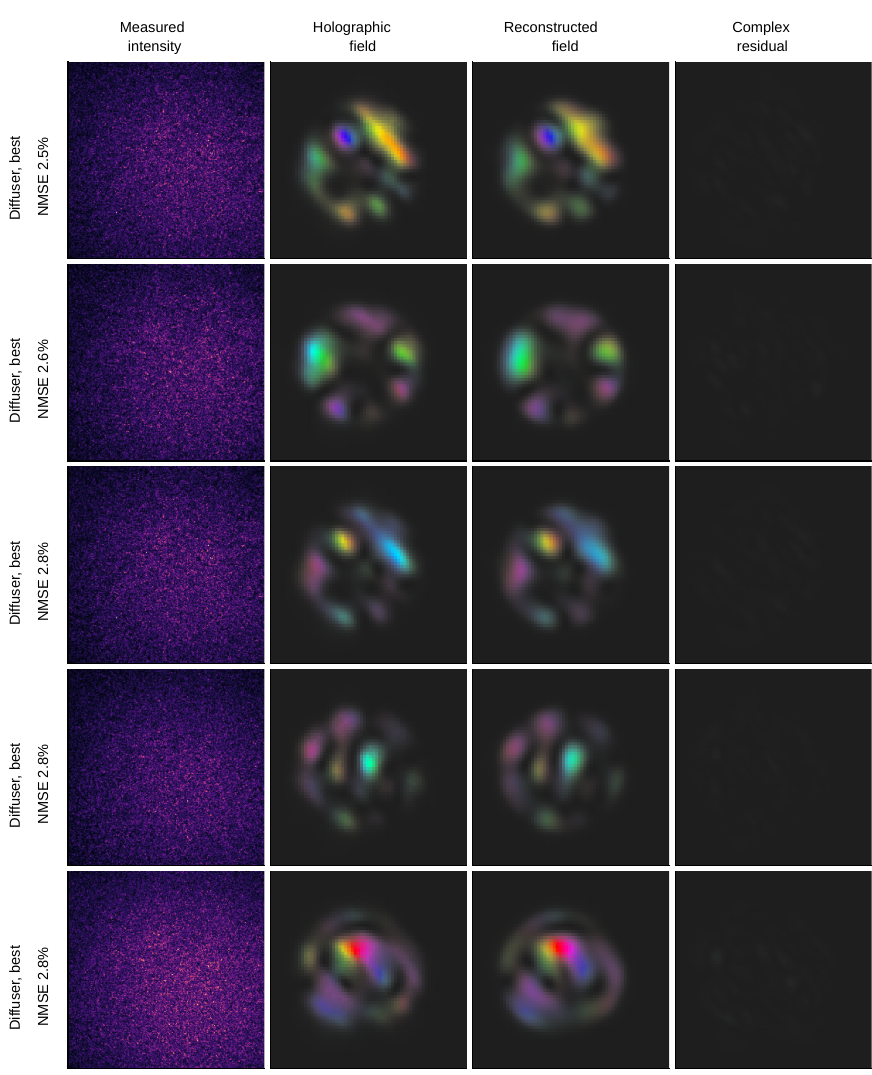}
\caption{Five eligible diffuser reconstructions with the lowest field NMSE. Columns show measured intensity, holographic reference, reconstructed field and complex residual. Reconstructions are aligned by a global complex scalar; reference, reconstruction and residual share the reference intensity scale within each row. Hue encodes phase and brightness/saturation intensity.}
\label{fig:S-diffuser-gallery}
\end{figure}
\clearpage
\begin{figure}[p]
\centering
\includegraphics[width=0.9\linewidth]{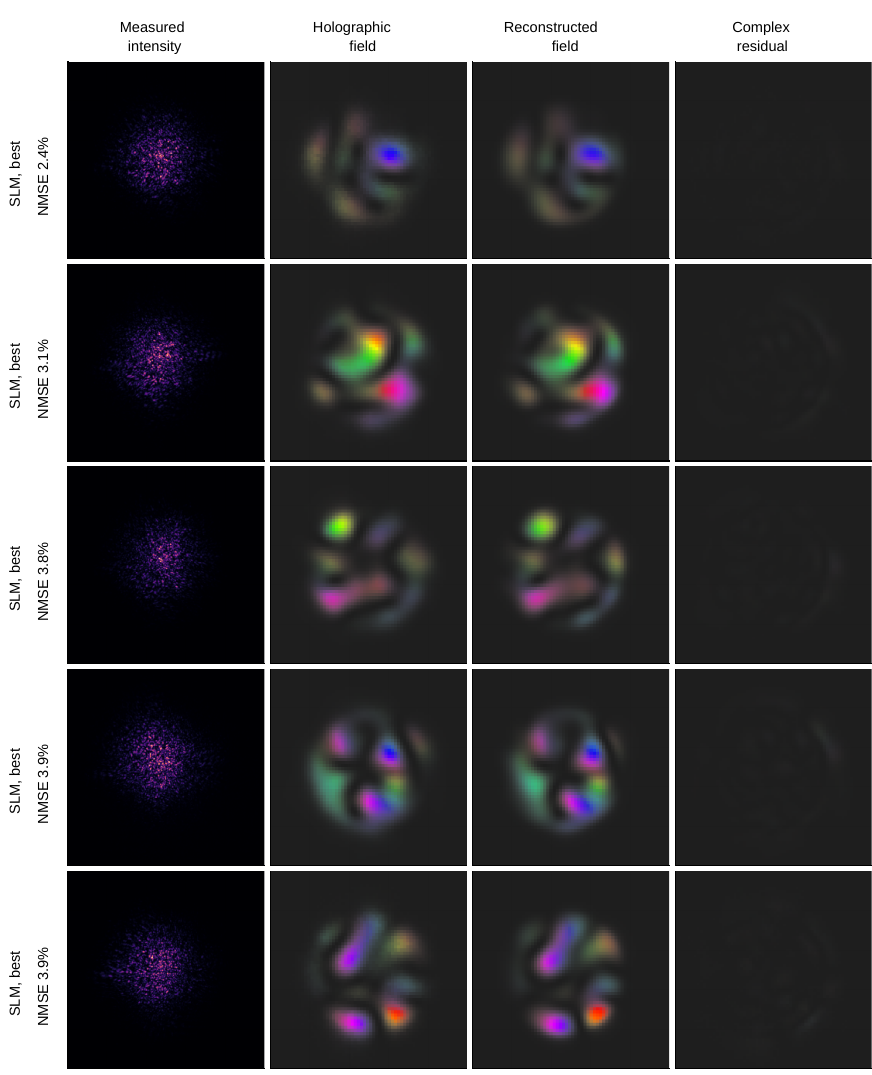}
\caption{Five eligible SLM reconstructions with the lowest field NMSE. Columns show measured intensity, holographic reference, reconstructed field and complex residual. Reconstructions are aligned by a global complex scalar; reference, reconstruction and residual share the reference intensity scale within each row. Hue encodes phase and brightness/saturation intensity.}
\label{fig:S-slm-gallery}
\end{figure}
\clearpage

\begin{figure}[p]
\centering
\includegraphics[width=0.9\linewidth]{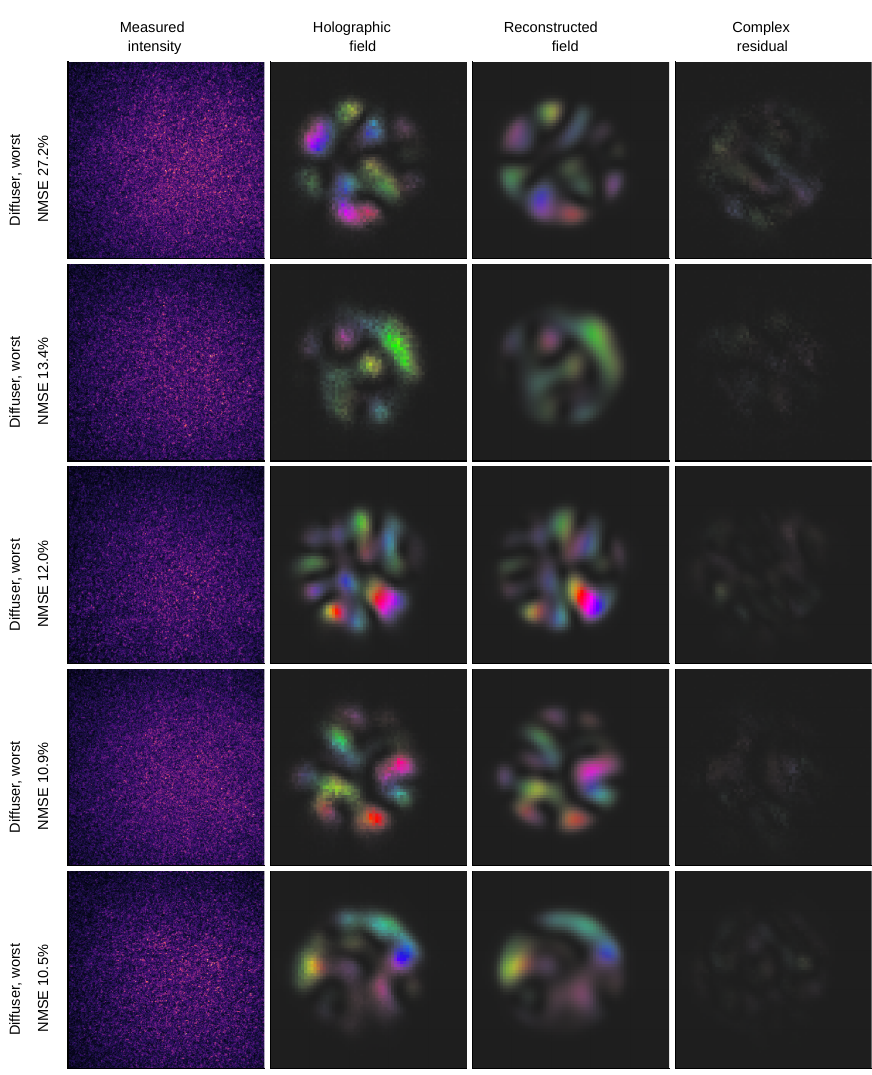}
\caption{Five eligible diffuser reconstructions with the highest field NMSE. Columns show measured intensity, holographic reference, reconstructed field and complex residual. Reconstructions are aligned by a global complex scalar; reference, reconstruction and residual share the reference intensity scale within each row. Hue encodes phase and brightness/saturation intensity.}
\label{fig:S-diffuser-worst}
\end{figure}
\clearpage
\begin{figure}[p]
\centering
\includegraphics[width=0.9\linewidth]{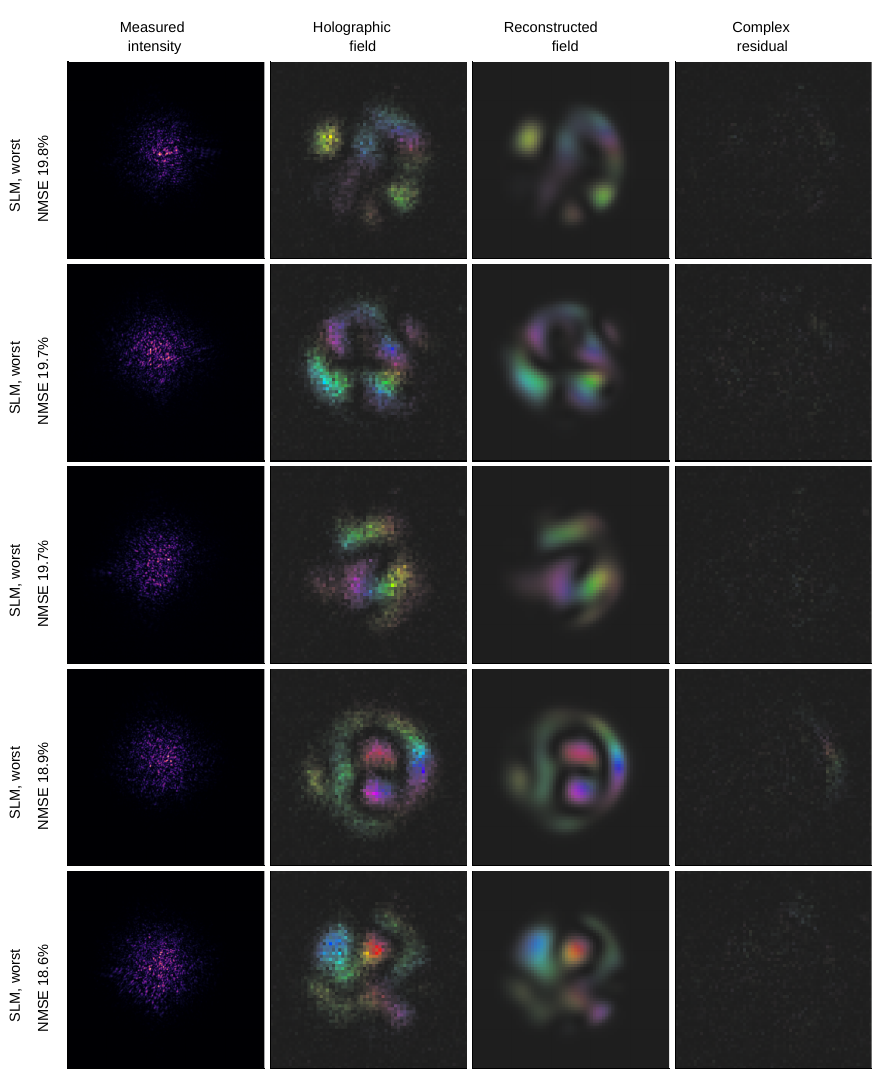}
\caption{Five eligible SLM reconstructions with the highest field NMSE. Columns show measured intensity, holographic reference, reconstructed field and complex residual. Reconstructions are aligned by a global complex scalar; reference, reconstruction and residual share the reference intensity scale within each row. Hue encodes phase and brightness/saturation intensity. Relative to the LP projection of each reference, their errors are 0.7--1.2\%; most of the discrepancy lies outside the LP subspace of the holographic references (out-of-subspace fractions 18--19\%).}
\label{fig:S-slm-worst}
\end{figure}
\clearpage

\subsection{Near-field control}

The eligible test fields were reconstructed from near-field intensities using the LP synthesis fields of the calibrated operator as near-field basis, the same ridge rule $\lambda_{\mathrm r}=10^{-3}\operatorname{tr}(G)/1,764$ and rank-one projection (\texttt{reporting/near\_field\_control.py}). The near-field PMP matrix keeps 286 of 1,764 directions above the single-precision cutoff $\lambda_j>2^{-23}\lambda_{\max}$, against all 1,764 for the calibrated symmetry-broken-plane operators. For the diffuser acquisition, the reference-blocked near-field images on camera C2 give a median field NMSE of 94.0\% (368 fields). Near-field intensities computed from the holographic reference fields, $|E_{\mathrm{holo}}|^2$, which are free of calibration and sampling mismatch and are the only near-field data of the SLM acquisition, give 96.4\% (diffuser) and 98.2\% (SLM, 150 fields), against 4.6\% and 9.1\% from the symmetry-broken-plane images. Allowing for complex conjugation of the field does not reduce these errors, so the failure is not only the conjugate ambiguity.

\begin{figure}[!ht]
\centering
\includegraphics[width=0.48\linewidth]{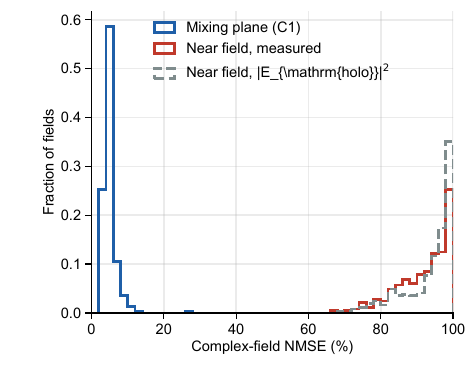}\hfill
\includegraphics[width=0.48\linewidth]{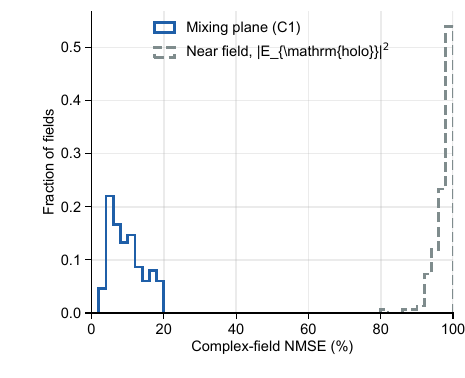}\\[4pt]
\includegraphics[width=0.48\linewidth]{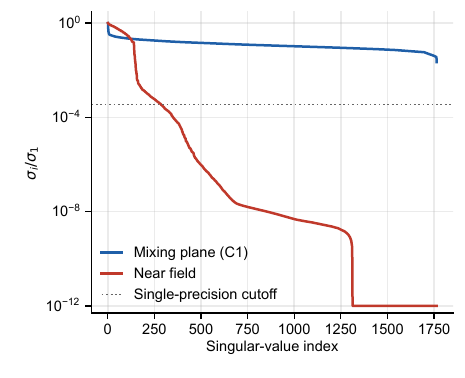}
\caption{Near-field control. Top: distributions of the complex-field NMSE of the same eligible test fields reconstructed from the symmetry-broken-plane image (C1), from the measured near-field image (C2; diffuser acquisition only) and from the near-field intensity of the holographic field, for the diffuser (left) and SLM (right) acquisitions; bin width 2\%. Bottom: normalised singular values of the calibrated symmetry-broken-plane PMP matrix of the SLM acquisition and of the near-field PMP matrix; the dotted line marks the single-precision cutoff, and numerical zeros are drawn at $10^{-12}$.}
\label{fig:S-near-field}
\end{figure}

\subsection{Data and code availability}
The data and code supporting this study will be archived in a Zenodo record (DOI: \url{https://doi.org/10.5281/zenodo.23785123}). The record will contain the complete raw diffuser and optimised-SLM acquisitions as separate HDF5 files, together with the processed-data and code archive. The archive includes historical processed calibration inputs, frozen operators, processed held-out measurements, saved reconstructions and coefficients, eligibility masks, inference configurations, validation records, figure inputs and source snapshots. The manifest records file identities and SHA-256 checksums.

The principal replay and calibration entry points are
\begin{itemize}
\item \path{scripts/run_replay_checks.py};
\item \path{scripts/calibration/run_cuda_verification.py}.
\end{itemize}
The files \path{README.md}, \path{INSTALL.md} and \path{FIGURE_TABLE_MAP.md} give the commands and tested environments. Replaying all 500 held-out symmetry-broken-plane images for each acquisition reproduces the saved reconstructions to aligned NMSE $2.4\times10^{-8}$ (diffuser) and $1.5\times10^{-10}$ (SLM), and reproduces the reported eligible metrics. Replaying the frozen operators on raw frames of the ten displayed examples gives aligned NMSE $1.7\times10^{-8}$ (diffuser) and $7.0\times10^{-11}$ (SLM). The complete raw acquisitions are included separately in the Zenodo record, allowing the full measurement sets to be inspected and reprocessed.

The frozen calibration is reproduced bitwise from the included historical processed arrays. Recalibration from raw data reproduces the SLM result exactly. For the diffuser, float32 hologram-extraction differences of order $10^{-6}$ lead to a refitted eligible median NMSE of 4.6129\% rather than 4.6051\% from the historical arrays; this raw-derived refit exceeds the strict frozen-operator comparison tolerance. The release \path{SCOPE.md} states the reproducibility limits and comparison tolerances. The three warm-start checkpoints used for calibration are included, while their earlier training lineage is outside the scope of this release. Large simulation campaigns and the Apple Metal benchmark are provided as saved results with source/configuration records rather than rerun here. The original saved stochastic camera-intensity arrays used for all eight SI median-example panels are included and are read directly by the figure builder. The two acquisitions do not isolate a causal benefit from optical optimisation or condition number.

\clearpage
\section{Supplementary Note 7: Other modal families}

We applied the same optical geometry and optimisation procedure to two additional modal families. All modes are sampled on the fibre-plane grid of the LP study ($256\times256$ pixels over $\pm25.16\,\mu$m), normalised to unit power and propagated through the same simulated optical system to the same $1024\times1024$ phase plane and camera, with the same soft ROI.

\emph{Hermite--Gaussian beams.} HG$_{mn}$ with $m+n\le7$ (36 modes) and waist $w_0=5.5\,\mu$m, chosen so that the far-field divergence of the highest order matches the 0.10 numerical aperture of the 42-mode fibre. \emph{Square waveguide.} Scalar guided modes of a square step-index core ($n_{\mathrm{clad}}=1.46$, numerical aperture 0.10, source wavelength 632.8 nm, rounded to 633 nm in the numerical model) from a finite-difference solution of the scalar Helmholtz equation; the core half-width of $11.25\,\mu$m ($V=11.17$) is the largest that guides at most 42 modes and gives 41 modes, including 10 degenerate pairs related by the exchange of $x$ and $y$. \emph{Reference.} The 42 LP modes of the paper with the seed-12 mask, evaluated by the same code.

For each new family the protocol comprised: the near and far fields; a flat-phase SLM; random pixel masks of cell size 1, 4, 16 and 32 pixels (four each); the fibre-optimised seed-12 mask transferred unchanged; and the ROI-constrained optimisation of Eq.~(5) of the main text with the optimisation schedule of Supplementary Table~\ref{tab:S-optimization-settings}, from three initial phases (seeds 12, 23 and 34). Condition numbers use all camera pixels and all $M^2$ products. Reconstruction errors use 200 random fields per SNR, the 10-bit camera model of Supplementary Note~5, the unregularised full-camera pseudoinverse and rank-one projection. Robustness is the condition number of the seed-12 optimised mask after shifting it by 1--16 pixels (8--128\,$\mu$m) along one axis or the diagonal, or changing the mask--camera distance by up to $\pm2$ mm. The calculations ran on an Apple-silicon GPU (Metal), with Gram matrices accumulated in double precision; the condition number of the paper's mask for the LP modes is reproduced (19.97).

\begin{figure}[!ht]
\centering
\includegraphics[width=\linewidth]{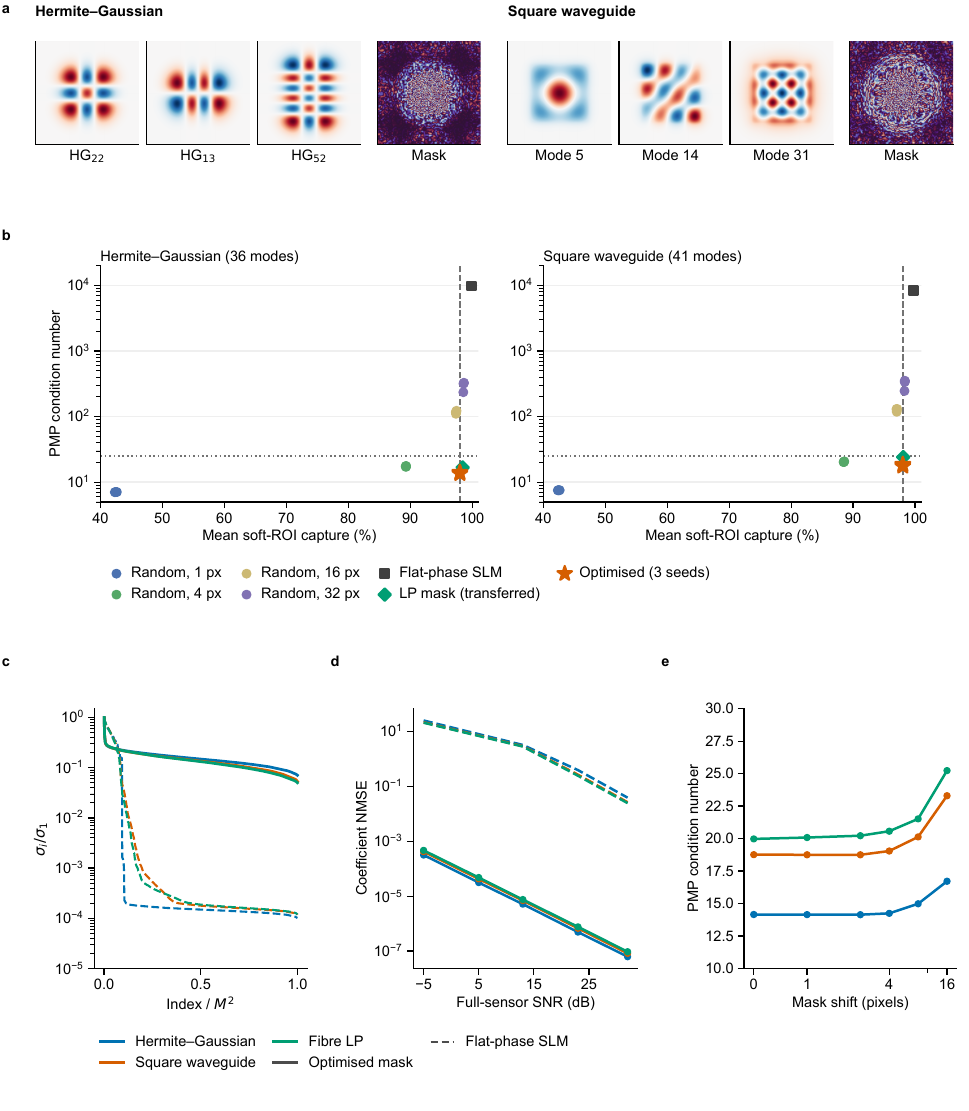}
\caption{The measurement-design procedure applied unchanged to other modal families (simulation). \textbf{(a)} Three of the 36 Hermite--Gaussian modes and three of the 41 square-waveguide modes (real fields), with the central $5.1\times5.1$ mm of each family's optimised mask (seed 12; zero phase dark). \textbf{(b)} Full-camera PMP condition number versus mean soft-ROI capture for random masks, a flat-phase SLM, the fibre-optimised mask transferred unchanged and the optimised masks of three initial phases; dashed and dotted lines mark the 98\% capture and $\kappa=25$ targets. \textbf{(c)} Normalised singular values of the optimised (solid) and flat-phase-SLM (dashed) PMP matrices, against index divided by $M^2$; values below $10^{-5}$ are off scale. \textbf{(d)} Median coefficient NMSE of single-image reconstruction versus full-sensor SNR, 200 fields per point. \textbf{(e)} Condition number of each family's optimised mask (fibre: the mask of the paper) versus transverse mask shift along one axis.}
\label{fig:S-universality}
\end{figure}

Supplementary Table~\ref{tab:S-universality} and Supplementary Fig.~\ref{fig:S-universality} summarise the results. In all three families the near- and far-field PMP matrices are singular, the flat-phase SLM is nearly singular, and no random mask satisfies both capture constraints with $\kappa\le25$: fine random masks condition the products but lose most of the light, and coarse ones keep the light but condition poorly. For the two additional families, optimised masks satisfy both constraints with $\kappa$ between 13.5 and 18.8, and the reconstruction error then scales as 1/SNR in the same way as for the fibre. The transferred fibre mask already satisfies the constraints for both new families, at a higher $\kappa$, showing that a mask optimised for one basis breaks much of the symmetry of others; re-optimisation for each basis lowers $\kappa$ by 16\% (Hermite--Gaussian) and 22\% (waveguide). In all families the condition number rises by at most 12\% for mask shifts up to 8 pixels (64\,$\mu$m) and by at most 6\% for distance errors of $\pm2$ mm. These are matched-model simulations: the same optical model generates and inverts the data, and the waveguide modes are scalar.

\begin{table}[!ht]
\centering
\caption{Reconstruction for additional modal families (simulation). $\kappa$: full-camera PMP condition number. Capture: mean soft-ROI / worst coherent state (\%). NMSE: median coefficient NMSE at 13 dB. Optimised: seed 12 (seeds 23 and 34 in brackets).}
\label{tab:S-universality}
\small
\begin{tabular}{lccc}
\toprule
 & Hermite--Gaussian & Square waveguide & Fibre LP (paper)\\
\midrule
Modes $M$ & 36 & 41 & 42\\
Near-field rank / $M^2$ & 120 / 1,296 & 281 / 1,681 & 282 / 1,764\\
Flat-phase SLM $\kappa$; NMSE & 9,829; 3.3 & 8,446; 2.9 & 8,378; 2.9\\
Random, 1 px: median $\kappa$; capture & 7.1; 42 & 7.6; 42 & 7.6; 42\\
Random, 32 px: median $\kappa$; capture & 322; 98.6 & 343; 98.3 & 336; 98.3\\
Transferred fibre mask $\kappa$; capture & 16.8; 98.5 / 97.2 & 24.1; 98.1 / 96.4 & --\\
Optimised $\kappa$ & 14.2 (14.0, 13.5) & 18.8 (18.3, 17.5) & 20.0\\
Optimised capture & 98.0 / 95.8 & 98.0 / 96.0 & 98.0 / 96.4\\
Optimised NMSE & $5.1\times10^{-6}$ & $7.0\times10^{-6}$ & $7.5\times10^{-6}$\\
$\kappa$, shift 8 / 16 px & 15.0 / 16.7 & 20.1 / 23.3 & 21.5 / 25.2\\
$\kappa$, $\Delta z=\pm2$ mm & 13.9--14.8 & 18.3--19.6 & 19.3--21.0\\
\bottomrule
\end{tabular}
\end{table}

\end{document}